\documentclass[%
preprint,
 amsmath,amssymb,
 aps,
floatfix,
]{revtex4-1}
\usepackage{gensymb}
\usepackage{graphicx}% Include figure files
\usepackage{dcolumn}% Align table columns on decimal point
\usepackage{bm}% bold math
\usepackage{booktabs}
\usepackage{color}

\usepackage{hyperref}
\hypersetup{
    bookmarks=true,         % show bookmarks bar?
    unicode=false,          % non-Latin characters in Acrobat’s bookmarks
    pdftoolbar=true,        % show Acrobat’s toolbar?
    pdfmenubar=true,        % show Acrobat’s menu?
    pdffitwindow=false,     % window fit to page when opened
    pdftitle={Fe3Sn},    % title
    pdfauthor={Dr. Olga Vekilova},     % author
    pdfsubject={magnetic},   % subject of the document
    pdfcreator={Dr. Olga Vekilova},   % creator of the document
    pdfproducer={},  % producer of the document
    pdfkeywords={Fe3Sn} {anisotropy} {magnetic}, % list of keywords
    pdfnewwindow=true,      % links in new window
    colorlinks=false,       % false: boxed links; true: colored links
    linkcolor=red,          % color of internal links
    citecolor=green,        % color of links to bibliography
    filecolor=magenta,      % color of file links
    urlcolor=cyan           % color of external links
}

\begin{document}

%\preprint{APS/123-QED}

\title{Tuning magnetocrystalline anisotropy of $\text{Fe}_{3}\text{Sn}$ by alloying}% Force line breaks with \\

\author{Olga Yu. Vekilova}
 \email{olga.vekilova@physics.uu.se}
\affiliation{Department of Physics and Astronomy, Uppsala University, Box 516, 75121 Uppsala, Sweden}%


\author{Bahar Fayyazi}
\affiliation {Materials Science, TU Darmstadt, Alarich-Weiss-Str. 16, 64287 Darmstadt, Germany}

\author{Konstantin P. Skokov}
\affiliation {Materials Science, TU Darmstadt, Alarich-Weiss-Str. 16, 64287 Darmstadt, Germany}

\author {Oliver Gutfleisch}
\affiliation {Materials Science, TU Darmstadt, Alarich-Weiss-Str. 16, 64287 Darmstadt, Germany}

\author{Cristina Echevarria-Bonet} 
\affiliation {BCMaterials, UPV/EHU Science Park, 48940 Leioa, Spain}

\author{Jos\'e Manuel Barandiar\'an}
\affiliation {BCMaterials, UPV/EHU Science Park, 48940 Leioa, Spain}

\author{Alexander Kovacs}
\affiliation{Department for Integrated Sensor Systems, Danube University Krems, Viktor Kaplan Str. 2/E, 2700 Wiener Neustadt, Austria}

\author{Johann Fischbacher}
\affiliation{Department for Integrated Sensor Systems, Danube University Krems, Viktor Kaplan Str. 2/E, 2700 Wiener Neustadt, Austria}

\author{Thomas Schrefl}
\affiliation{Department for Integrated Sensor Systems, Danube University Krems, Viktor Kaplan Str. 2/E, 2700 Wiener Neustadt, Austria}

\author{Olle Eriksson}%
\affiliation{Department of Physics and Astronomy, Uppsala University, Box 516, 75121 Uppsala, Sweden}%
\affiliation{School of Science and Technology, \"Orebro University, SE-701 82 \"Orebro, Sweden}

\author{Heike C. Herper}
\affiliation{Department of Physics and Astronomy, Uppsala University, Box 516, 75121 Uppsala, Sweden }%
%\collaboration{NOVAMAG Collaboration}
%\affiliation{
%\affiliation{
 %Third institution, the second for Charlie Author
%\affiliation{%


\date{\today}% It is always \today, today,

\begin{abstract}
The electronic structure, magnetic properties and phase formation of hexagonal ferromagnetic Fe$_3$Sn-based alloys have been studied from first principles and by experiment. % by means of highly accurate full-potential linear muffin-tin orbital (LMTO) method implemented in the RSPt code. 
The pristine Fe$_3$Sn compound is known to fulfill all the requirements for a good permanent magnet, except for the magnetocrystalline anisotropy energy (MAE). The latter is large, but planar, i.e. the easy magnetization axis is not along the hexagonal c direction, whereas a good permanent magnet requires the MAE to be uniaxial. Here we consider Fe$_3$Sn$_{0.75}$M$_{0.25}$, where M= Si, P, Ga, Ge, As, Se, In, Sb, Te and Bi, and show how different dopants on the Sn sublattice affect the MAE and can alter it from planar to uniaxial. %The calculation of exchange iterations allows us to predict the Curie temperature (T$_C$) of the system within the mean-field approximation, as well as by means of a Monte Carlo simulation. 
The stability of the doped Fe$_3$Sn phases is elucidated theoretically via the calculations of their formation enthalpies. A micromagnetic model is developed in order to estimate the energy density product $(BH)_\mathrm{max}$ and coercive field $\mu_{0}H_\mathrm{c}$ of a potential magnet made of Fe$_{3}$Sn$_{0.75}$Sb$_{0.25}$, the most promising candidate from theoretical studies. 
The phase stability and magnetic properties of the Fe$_3$Sn compound doped with Sb and Mn has been checked experimentally on the samples synthesised using the reactive crucible melting technique as well as by solid state reaction. The Fe$_3$Sn-Sb compound is found to be stable when alloyed with Mn.  
It is shown that even small structural changes, such as a change of the c/a ratio or volume, that can be induced by, e.g., alloying with Mn, can influence anisotropy and reverse it from planar to uniaxial and back.
%The calculated MAE  have been analyzed and suggestions for the improvement of permanent magnets have been formulated.

\begin{description}
\item[PACS numbers] 75.50.Ww, 75.30.Gw, 75.20.En
\end{description}
\end{abstract}

\pacs{Valid PACS appear here}% PACS, the Physics and Astronomy
\maketitle

%\tableofcontents

\section{\label{sec:level1}Introduction}

Strong permanent magnets creating high magnetic field are of ultimate importance for many technological applications, from magnetic resonance imaging to magnetic hard disk drives in information storage.  They are also used in a number of green energy applications, like motors for hybrid and electric cars and direct-drive wind turbines \cite{Perm_magn_review}.
The strongest known permanent magnets typically contain rare earth elements \cite{Oliver}. For the past few decades the demand for such magnets has substantially increased. As  the cost of the rare-earth based materials is high, the search for magnets, that are cheaper and contain  smaller amounts of rare earth elements, has become an important field of research \cite{Perm_magn_review,Bahar_nn,Bahar_nn2}.

Ferromagnets with rather high Curie temperature (T$_C$) above 400 K and  high saturation magnetization as well as high magnetocrystalline anisotropy energy (MAE) are considered as good candidates for permanent magnet applications. Furthermore, for such materials the axis corresponding to the longest lattice constant, i.e. the c axis in most hexagonal structures, should be the unique easy magnetization direction \cite{Perm_magn_review,Bahar_nn,Bahar_nn2}. These properties can be found in particular in Fe-rich materials with non-cubic uniaxial crystal structures. The hexagonal Fe$_3$Sn compound satisfies these conditions to a large degree. Among five existing intermetallic compounds containing Fe and Sn the Fe$_3$Sn phase is one of the most attractive ones, due to the highest concentration of iron and therefore highest magnetic moment. The other advantages of the rare-earth free Fe$_3$Sn system are its relatively low price and rather high T$_C$ of about 743 K \cite{T_C_exp}.
However, as it has recently been shown both experimentally and theoretically, the magnetocrystalline anisotropy of Fe$_3$Sn is planar, which is undesirable for a permanent magnet \cite{Sales_Fe3Sn}. It has also been suggested by Sales et. al that alloying with Sb might change the anisotropy to uniaxial \cite{Sales_Fe3Sn}. 
Motivated by this research we studied the influence of different dopants on the MAE of the Fe$_3$Sn compound. The electronic structure and magnetic properties, as well as the effect of the hexagonal c/a ratio on the magnetocrystalline anisotropy in the Fe$_3$Sn compound and its alloys were addressed from first principles by means of highly accurate full-potential linear muffin-tin orbital (LMTO) method implemented in the RSPt code. The stability of the doped phases were elucidated by the calculation of formation enthalpies by means of the VASP code.

Theoretical calculations were combined with an experimental study in order to verify the phase stability as well as the magnetic properties of Fe$_3$Sn compounds. Combinatorial materials science and high-throughput screening methods, such as Reactive Crucible Melting (RCM) technique, offer an efficient strategy for the discovery of new materials with promising properties. Using the knowledge of the required synthesis conditions for the formation of the Fe$_3$Sn phase \cite{Bahar_10}, we performed the RCM method in order to search for the (Fe)$_3$(SnM) phase. To get more insight into the magnetic properties of the (Fe)$_3$(SnM) phase, additional experiments were performed using the Solid State Reaction (SSR) method, which was used for the preparation of the desirable compound from the mixture of the starting elements by means of atomic diffusion. 
For the most promising system a micromagnetic model was developed to calculate the magnetic induction as a function of the internal field.  
%\color{red}(*) \color{black}

The paper is organized as follows. Section \ref{sec:theory} describes theoretical methods used for the first-principles calculations. Section \ref{sec:exp} provides the experimental details. In sections \ref{sec:theor_stab}, \ref{sec:theor_anis}, \ref{sec:exp_res} and \ref{sec:exp_res_c} the results of theoretical simulations as well as experimental results are given. Section \ref{sec:micromagn} addresses the results of the micromagnetic simulations. Conclusions are given in section \ref{sec:conclusions}.

%These approaches can rapidly create material libraries as multiple samples are simultaneously synthesized and subsequently their properties are characterized by high-throughput methods.

%Several phase diagrams of material systems have been constructed using different combinatorial approaches \cite{B1,B2}  e.g. thin-film depositions techniques \cite{B3} and bulk high-throughput techniques such as reactive diffusion methods \cite{B4} and reactive crucible melting \cite{B5,B6,B9}.%[5-9].

\section{\label{sec:level1}Methods}

\subsection{\label{sec:theory}Theory}

\begin{figure} [tbp] 
%\begin{center}  
\includegraphics[scale=0.1]{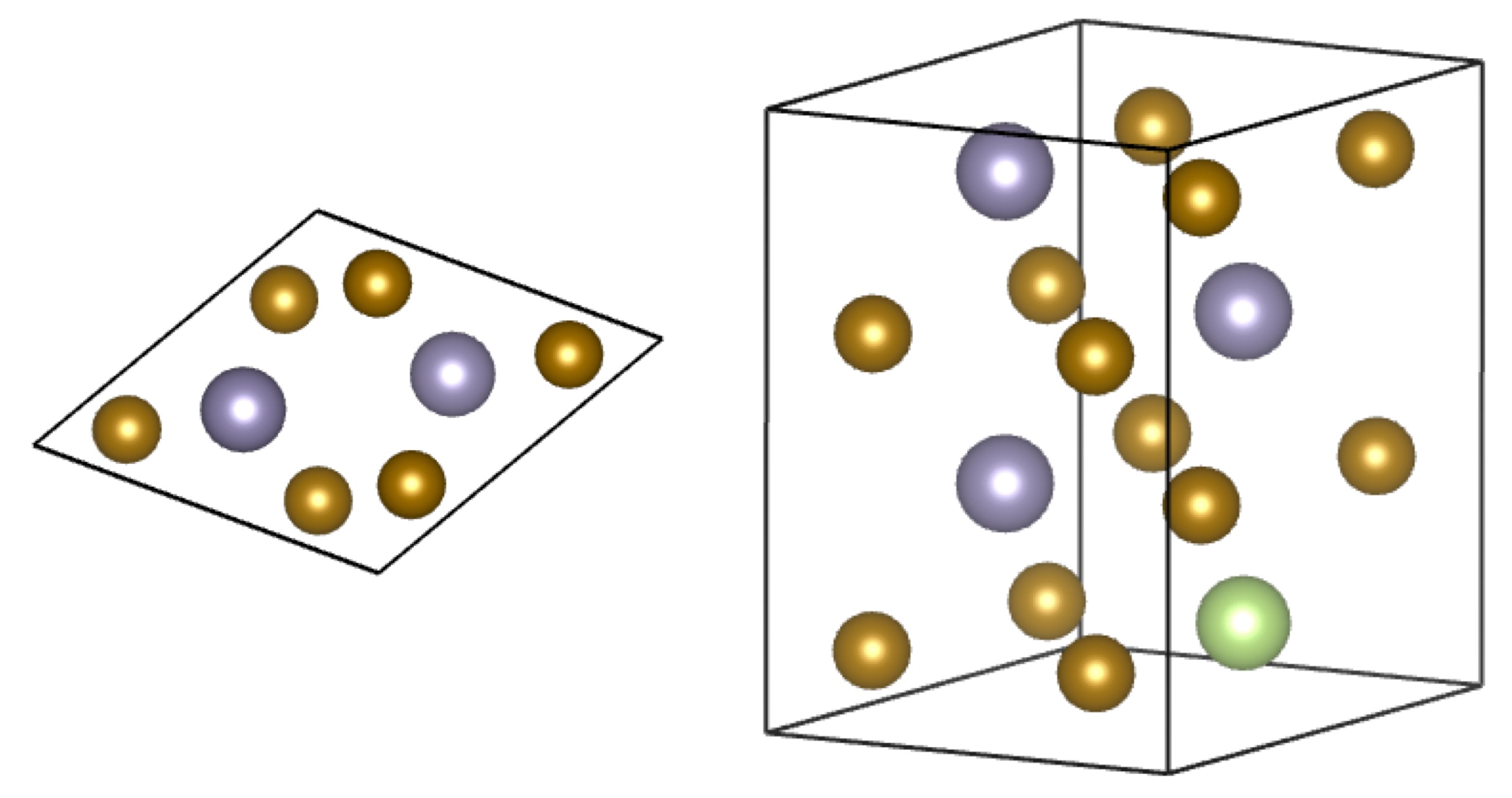} 
%\end{center} 
\caption{(Color online) $1\times1\times2$ Fe$_3$Sn hexagonal cell with one impurity atom on the tin sublattice. Iron atoms are shown with brown spheres, Sn atoms with grey and M (M=Si, P, Ga, Ge, As, In, Sb, Te and Bi) impurity atom is shown with the green sphere.}
\label{fig:struc}  
\end{figure}

 The high temperature phase of Fe$_3$Sn  has a hexagonal crystal structure with space group P6$_3$/mmc ($\#194$) and contains 8 atoms per unit cell (see Fig. \ref{fig:struc}). For calculations of phase stability Vienna Ab Initio Simulation Package (VASP) \cite{VASP_1, VASP_2, VASP_3} was used within the projector augmented wave (PAW) method \cite{PAW}. The electronic exchange and correlation ef\mbox{}fects were treated by the generalized gradient approximation (GGA) in the Perdew, Burke, and Ernzerhof (PBE) form \cite{GGA} in all the used methods. A 64 atom supercell of Fe$_3$Sn comprising of 46  Fe and 16 Sn atoms was considered. The plane-wave energy cut-off was set to 350 eV. The converged k-point mesh was found to be $8\times8\times8$ k-points.  The obtained magnetic moment on iron  was $\sim 2.4 \mu_{B}$ per atom, in agreement with previous studies \cite{Sales_Fe3Sn}. 
 
 The Sn atoms of the Fe$_3$Sn compound were partially replaced by alloying elements. In order to examine the phase stability of considered doped systems for each impurity, i.e. M=Sb, Ga, Ge, and Hf,  we calculated 3 different distributions of impurity atoms, namely ordered, random (i.e. mimicking a disordered alloy), and phase separated (i.e. mimicking clusterization of dopants) (See Supplemental Materials). We used the special quasi-random structure (SQS) technique \cite{SQS,Abrikosov_SQS,Ruban_SQS} to generate the corresponding supercells. %Also, for each of the considered phases we studied two magnetic distributions of iron atoms, namely the ferromagnetic (fm) order and the paramagnetic state in the framework of the disordered local moment (dlm) model \cite{Pindor1983}.  
For the estimation of phase stability of ternary compounds the following equation was used:
 \begin{equation}
\Delta H=H_{\mathrm{Fe}_3\mathrm{Sn}_x\mathrm{M}_{1-x}} - x H_{\mathrm{Fe_3Sn}} - (1-x) H_{\mathrm{Fe_3M}}
\label{eq:stab}
\end{equation}
where $H_{\mathrm{Fe_3Sn}_x\mathrm{M}_{1-x}}$ is the enthalpy of a ternary compound and $1-x$ is the concentration of M, an impurity element on the Sn sublattice. 

 The $1\times1\times2$ supercell of Fe$_3$Sn comprising of 12 iron and 4 tin atoms, one of which is further substituted by an impurity (see Fig.  \ref{fig:struc}), was used for the calculation of the magnetic properties with help of the full-potential linear muf\mbox{}fin-tin orbital (FP-LMTO) method implemented in the RSPt code \cite{RSPT_1, RSPT_2}.  We performed integration over the Brillouin zone, using the tetrahedron method, with Bl\"{o}chl's correction \cite{bloechl_tetra}.   The k-point convergence of the MAE for the chosen supercell size was found when increasing the Monkhorst-Pack mesh \cite{monkhorst-pack} to $24\times24\times24$, that was further used in all calculations.
 
 The following impurities were considered in Fe$_3$Sn$_{0.75}$M$_{0.25}$ compound:  M=Si, P, Ga, Ge, As, Se, In, Sb, Te, and Bi. With one substitutional impurity atom in the considered supercell, the dopant concentration is fixed to 6.25 at.\%. The equilibrium lattice parameter and hexagonal c/a ratio were calculated for every structure using the VASP code.
The effective exchange interaction parameters (J$_{ij}$) were obtained using Lichtenstein et al. method \cite{Jij_1, Jij_2}, as implemented in RSPt \cite{Jij_3}. In this technique the energy of the system is mapped onto a classical Heisenberg model with the following Hamiltonian:
\begin{equation}
\hat{H}=-\frac{1}{2}\sum_{i\neq j}J_{ij}\vec{e}_i\cdot \vec{e}_j,
\label{eq:ham}
\end{equation}
where $\vec{e}_i$ denotes the unit vector along the magnetic moment at the site $i$. The exchange parameter between sites $i$ and $j$ is defined in the following way:
\begin{equation}
J_{ij}=\frac{T}{4}\sum_{n}Tr\Bigg[\hat{\Delta}_i(i\omega_n)\hat{G}^\uparrow_{ij}(i\omega_n)\hat{\Delta}_j(i\omega_n)\hat{G}^\downarrow_{ij}(i\omega_n)\Bigg],
\label{eq:jij}
\end{equation}
where T is the temperature, $\Delta$ is the on-site exchange potential, $G_{ij}$ is an inter-site Green’s function and $i\omega_n$ is the $n$-th fermonic Matsubara frequency \cite{Jij_3}. 
%The Curie temperature in the system can be roughly estimated within the mean-field approach, where it is proportional to the on-site exchange parameter J$_0$ using the following equation:
%\begin{equation}
%T_C=2J_0/3
%\label{eq:temp}
%\end{equation}
%A more accurate way to estimate the Curie temperature is to make use of Monte Carlo simulations. We used its implementation in the UppASD code \cite{UppASD_1, UppASD_2}.

 %Different concentrations of impurity atoms on the Sn sublattice, namely  6.25, 12.5, 18.75 and 25 at. \%,  were considered for the calculation of formation enthalpy (see Fig. \ref{fig:stab}).  On the left-hand side of Fig. \ref{fig:stab} the energies of these systems were compared with the total energy of the original Fe$_3$Sn system and on the right-hand side with the total energy of Fe$_3$M phase, when existing, or the sum of the energies of corresponding pure elements. 

 \begin{figure} [tbp] 
%\begin{center}  
\includegraphics[scale=0.12]{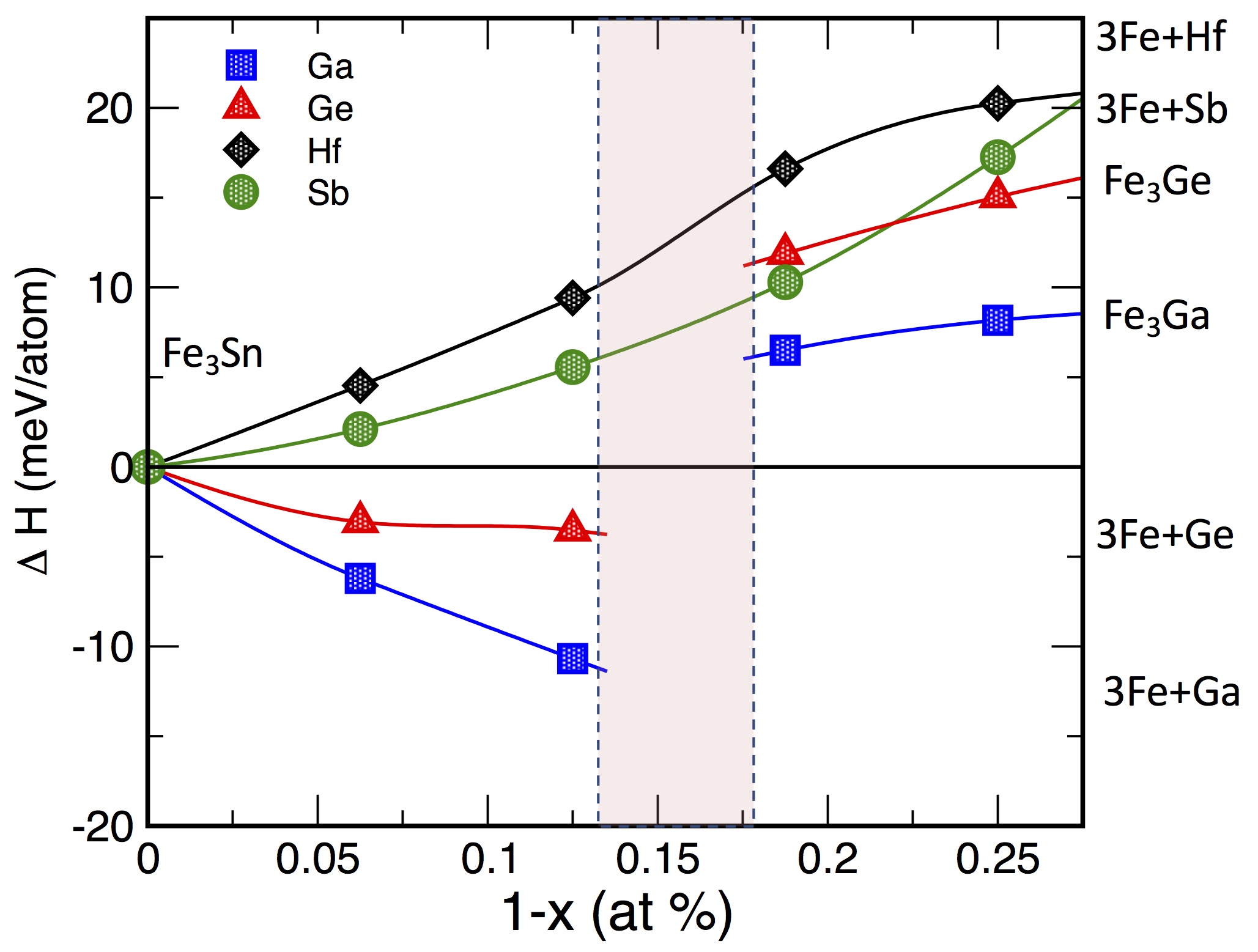} 
%\end{center} 
\caption{(Color online) Enthalpy of formation ($\Delta$H) of the ternary Fe$_3$Sn$_{0.75}$M$_{0.25}$ compound as a function of the concentration of dopant M, calculated in comparison with the Fe$_3$Sn binary on the left-hand side and Fe$_3$M (M=Ga, Ge and Hf) binary compound or a mixture of Fe and the M element on the right-hand side of the dashed vertical line.}
\label{fig:stab}  
\end{figure}

\subsection{\label{sec:exp}Experiment}
Screening through a large variety of material compositions and possible temperature stability ranges using traditional equilibrium alloy approach, which is a one-alloy-at-a-time practice, is a laborious task. Combinatorial materials science and high-throughput screening methods
enable the synthesis of large number of samples at once and therefore speed up the discovery of the new materials. Furthermore, using high-throughput characterization methods, their properties can be efficiently determined.
%offer an efficient strategy for discovery of new materials with promising properties, the Reactive Crucible Melting technique. This approach allows one to rapidly create material libraries as multiple samples are synthesized simultaneously and their properties are characterized by high-throughput methods. 
Several phase diagrams of material systems have been constructed using different combinatorial approaches \cite{B1,B2},  e.g. thin film deposition \cite{B3} and bulk high-throughput techniques, such as reactive diffusion method \cite{B4}, and reactive crucible melting \cite{B5,B6,B9,Bahar_8,B_7,Bahar_10}.%[5-9]

The RCM method is firstly introduced in Ref. \cite{Bahar_8} as a tool to search for hard magnetic phases. The method is based on diffusion processes driven by the formation of concentration gradient between the crucible material and other elements which are filled into it. Production procedure and working principle of the RCM method is fully described in Ref. \cite{Bahar_10}, where the method is applied to the Fe-Sn binary system and all five known intermetallic compounds of the system were synthesized in the reactive crucibles. The Fe$_3$Sn (3:1) phase was obtained in RCM by quenching of the crucibles after annealing at high temperatures (1023~K-1098~K). %\degree C).
 
To explore the hexagonal Fe$_3$Sn phase in the quest for uniaxial anisotropy, Fe-Sn$_x$M$_{1-x}$ crucibles with M = Sb, Si, Ga, Ge, Pb, In, Bi and $0.5<x<0.75$ were synthesized. To further extend our search, quaternary Fe-Mn, Sn, Sb crucibles were additionally produced. The crucibles were made of 99.95 \% pure Fe and they were filled with about 1 g of the rest elements of the examined system. The filling elements with the purity $>99.9$ were added in the form of crushed pieces. The samples were annealed at three selected temperatures of 1013~K, 1043~K, and 1073~K %740\degree C, 770\degree C and 800\degree C
for one week and subsequently quenched. For details we refer to  \cite{Bahar_10}.

For high-throughput characterization, the microstructure of the formed phases was studied by Philips XL30 FEG scanning electron microscopy (SEM) in back-scattered electron (BSE) contrast mode and their chemical compositions were determined using energy dispersive x-ray (EDX) spectroscopy. In addition, a Zeiss Axio Imager.D2m magneto-optical Kerr effect (MOKE) microscopy was used to display the magnetic domain structure of the formed phases which may give a clear hint for identification of the phases with uniaxial anisotropy.%to the evaluation of the magnetic properties, e.g. MAE of the stabilized phases. 
 
%\color{red}SSR  Cristina
%\color{black}
Several Fe$_{y}$Mn$_{3-y}$Sn$_{x}$Sb$_{1-x}$ samples (y=3 and x=1; y=2.5, 2.25, 2, 1.5 and x=0.75; and y=1.5 and x = 0.9) were prepared by solid state reaction (SSR)  \citep{SSR_book} through the following procedure. First, stoichiometric amounts of powders of the starting elements (99.9+\% purity, particles of less than 50 microns in size) were handmilled with an agate mortar and pestle and then compacted into pellets using pressures up to 0.5 GPa at room temperature (RT). Pellets were then encapsulated in a quartz ampoule, heated to 1073~K %800\degree C
 for 48 hours in vacuum and then quenched into ice water. 
 
 This process was repeated twice in order to homogenize the composition in the sample.
X-Ray Diffraction (XRD) measurements were performed on a Philips X'Pert Pro diffractometer, in the Bragg-Brentano geometry, using Cu K$\alpha$ radiation ($\lambda = 1.5418$~\AA). The samples were placed on a spinner to avoid a possible preferential crystalline orientation. XRD patterns were analyzed by Rietveld refinements, through the FullProf Suite \cite{Cristina_1}, using a Thompson-Cox-Hastings pseudo-Voigt function to describe the profile of the peaks. Temperature dependent magnetization was measured up to 823~K%550\degree C
, using H=0.01~T, in a vibrating sample magnetometer EZ7-VSM from Microsense.

%This file may be formatted in either the \texttt{preprint} or
%\texttt{reprint} style. \texttt{reprint} format mimics final journal output. 
%Either format may be used for submission purposes. \texttt{letter} sized paper should
%be used when submitting to APS journals.

\section{\label{sec:results}Results}
\subsection{\label{sec:theor_stab} Phase stability}

The enthalpy of formation was calculated for ternary Fe$_3$Sn$_{0.75}$M$_{0.25}$ alloys, where M=Sb, Ga, Ge and Hf for different concentrations of dopants from 6.25, 12.5, 18.75 and up to 25 at. \% (see Fig. \ref{fig:stab}). The energies of these systems were examined and compared with the energies of the binary Fe$_3$Sn compound on the left-hand side and binary Fe$_3$M (or the mixture of pure Fe and M if the binary phase was not energetically favorable) on the right-hand side of Fig. \ref{fig:stab} (see also Eq. \ref{eq:stab}). For the sake of simplicity, whether this particular phase was compared with the binary phase or with a mixture of pure elements, is specified on the right hand side of Fig. \ref{fig:stab} for each considered structure. A positive slope corresponds to an unstable ternary compound, while the negative one corresponds to a stable structure (See Fig.\ref{fig:stab}). 

%The stability of structure doped with Ge and Ga was calculated relative to both mixture of pure elements and Fe$_3$M (M=Ga, Ge) phase. 
Investigating the phase diagrams of the Fe-Ga and Fe-Ge binaries \cite{Fe-Ga, Fe-Ge} one can see that in the region up to 10-20 at. \% of dopant elements at temperatures up to 1500~K the Fe$_3$M phase decomposes into a mixture of $\alpha$-Fe and pure Ga or Ge phases \cite{Fe-Ga, Fe-Ge}. At higher concentration of dopants the phase is, however, stable. %Taking this into account one can use this comparison when addressing low concentrations of dopants (lower than 15-20 at. \% depending on temperature), while at higher concentrations the structure should be compared with the existing binaries. 
For this reason, enthalpy of Fe$_3$Sn$_{0.75}$M$_{0.25}$ (M=Ga, Ge)  was calculated relative to both the mixture of pure elements as well as the Fe$_3$M (M=Ga, Ge) phase, see Fig. \ref{fig:stab}. The dashed vertical lines show the area around 15 at. \% of dopant concentration where the mixture of these phases might exist depending on temperature and the dopant. At a concentration of dopants lower than $\sim$ 15 at. \%, when compared to the mixture of pure elements on the right-hand side, the enthalpy is negative indicating stability of the structure (see lower red and blue lines in Fig. \ref{fig:stab}). However, when compared with the binary phase Fe$_3$M, the stability of the ternary structure should be estimated from the upper positive curves for both Ga and Ge dopants (see upper red and blue lines in Fig. \ref{fig:stab}). 
 %This is shown in Fig. \ref{fig:stab} with the two positive curves for Ga and Ge.  
 
 In the considered Fe$_3$Sn$_{0.75}$M$_{0.25}$ alloy the concentration of dopants is equal to 6.25 at \%, which means the structures could be considered as stable based on this theoretical estimation. In contrast, checking the phase diagram of Fe-Sb one can see that the binary Fe$_3$Sb phase cannot exist. Furthermore, when the formation enthalpy of Fe$_3$Sn$_{0.75}$Sb$_{0.25}$ phase was calculated in comparison with the mixture of pure elements on the right hand side (see the green curve in Fig. \ref{fig:stab}) it appeared that the phase is unstable. However, as it is shown in section \ref{sec:theor_anis}, this phase has shown promising magnetic properties, so in order to stabilize it experimentally Mn was added to the Fe sublattice, see sections \ref{sec:exp_res} and \ref{sec:exp_res_c}.

Similarly, for the Hf dopant even comparing with the mixture of elements (see the black line in Fig. \ref{fig:stab}), the enthalpy curvature is positive indicating thermodynamic instability of this phase. For that reason Hf was further excluded from consideration and calculation of the magnetic properties.

\subsection{\label{sec:theor_anis} Magnetocrystalline anisotropy}

We calculated the saturation magnetic moment of Fe$_3$Sn$_{0.75}$M$_{0.25}$ system. For the pristine Fe$_3$Sn it is equal to $\mu_0M_s=$ 1.49~T, or $M_s=$1.19~MA/m, in very good agreement with previous theoretical and experimental estimations \cite{Sales_Fe3Sn}. Due to the low concentration of the impurities, 6.25 at. \%,  in the doped system, the obtained value of saturation magnetization is close to the one of the undoped system. For instance, for the case of Sb impurity it is equal to 1.51~T, or 1.2~MA/m.

\begin{figure} [htbp] 
%\begin{center}  
\includegraphics[scale=0.12]{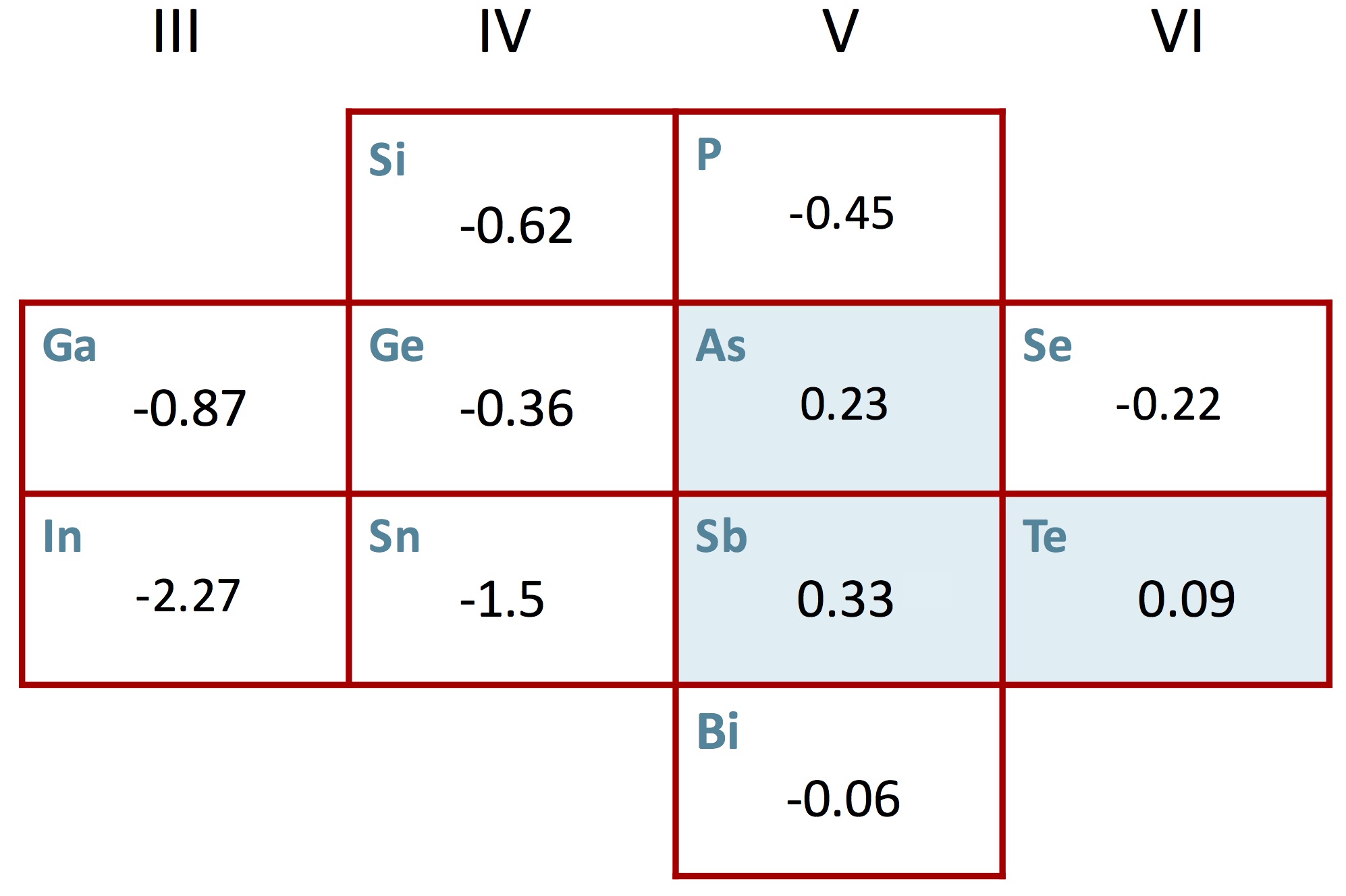} 
%\end{center} 
\caption{(Color online) Calculated values of magnetocrystalline anisotropy energy (MAE), K$_1$, and easy magnetization direction for different impurities for Fe$_3$Sn alloy grouped according to the positions of dopants in the Periodic Table.
 A negative value of K$_1$ indicates the in-plane easy magnetization axis, while the positive one shown with the filled blue rectangulars corresponds to the desirable uniaxial magnetocrystalline anisotropy.}
\label{fig:MAE_tab}  
\end{figure}

Magnetocrystalline anisotropy of the Fe$_3$Sn$_{0.75}$M$_{0.25}$ system, where M=Si, P, Ga, Ge, As, Se, In, Sb, Te and Bi was calculated from first principles electronic structure theory. The resulting MAE values are shown in Fig. \ref{fig:MAE_tab} for the dopants grouped according to their positions in the Periodic Table. For the pristine Fe$_3$Sn  the resulting anisotropy was found to be equal to -1.5 MJ/m$^3$ (see Fig. \ref{fig:MAE_tab}). In magnitude, this is one of the largest values among all the considered structures, however, the minus sign indicates the easy plane of magnetization. These results are in good agreement with other theoretical and experimental estimations (-1.59 MJ/m$^3$ and -1.8 MJ/m$^3$ respectively \cite{Sales_Fe3Sn}). 

 As one can see in Fig.  \ref{fig:MAE_tab}, the largest magnitude of the MAE (albeit favoring in-plane easy axis), of all the considered structures with dopants, was obtained for the Fe$_3$Sn$_{0.75}$In$_{0.25}$ compound. An uniaxial anisotropy was found for M=As, Sb and Te. However, the MAE for the Te doping is  very small. The anisotropy of  Fe$_3$Sn$_{0.75}$Sb$_{0.25}$ is in agreement with the existing theoretical data of 0.5  MJ/m$^3$ \cite{Sales_Fe3Sn}. As it was mentioned in Sec. \ref{sec:theor_stab}, in order to stabilize the Fe$_3$Sn$_{0.75}$Sb$_{0.25}$ system Mn was added on the Fe subblatice. The anisotropy of Fe$_{1.5}$Mn$_{1.5}$Sn$_{0.75}$Sb$_{0.25}$ system is equal to -1.49 MJ/m$^3$, which is very close to the value for the undoped  Fe$_3$Sn system. Adding Sb to the pristine Fe$_3$Sn allows for the change of anisotropy from planar to uniaxial, while adding Mn for stabilization reverts it back to planar.

\begin{figure}  [tbp] 
\includegraphics[scale=0.055]{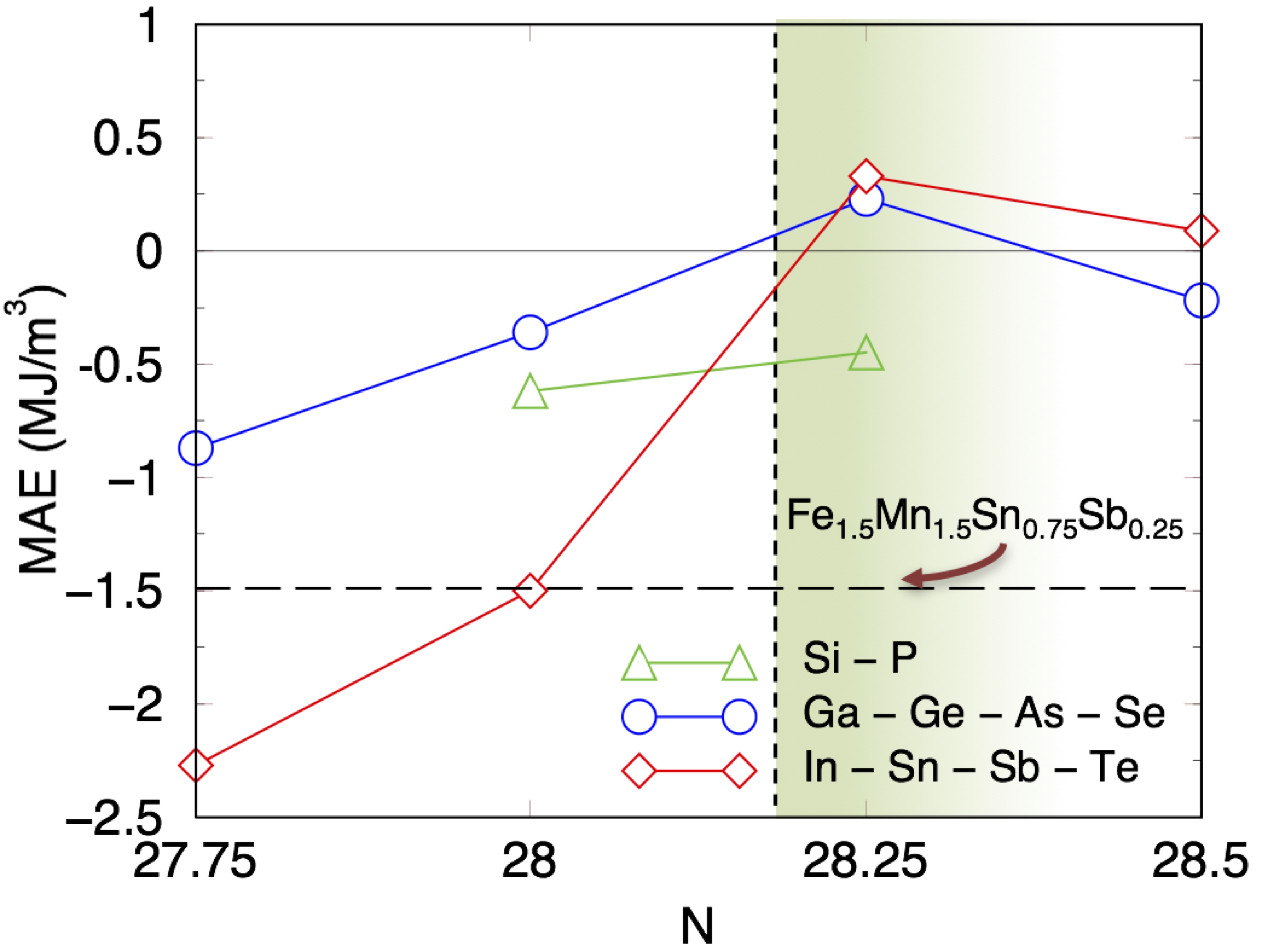} 
%\end{center} 
\caption{(Color online) The magnetocrystalline anisotropy energy (MAE) in Fe$_3$Sn$_{0.75}$M$_{0.25}$ compound as function of the number of valence electrons, N, of the system for three rows: Si-P (shown with the green line), Ga-Ge-As (shown with the red line) and In-Sn-Sb-Te (shown with the blue line). Doping from Group IIIA, Group IVA, Group VA and Group VIA, correspond to N=27.75, 28.25 and 28.5, respectively.The area of interest, where the easy magnetization axis is uniaxial, is shown with the gradient background. The horizontal dashed line shows the magnitude of the Fe$_3$Sn$_{0.75}$In$_{0.25}$ anisotropy, as its number of valence electrons is substantially lower.}
\label{fig:MAE_row}  
\end{figure}

The obtained MAE data were plotted as a function of the number of valence electrons per formula unit in the Fe$_3$Sn$_{0.75}$M$_{0.25}$ system for doping atoms from three rows of the Periodic Table, see Fig. \ref{fig:MAE_row}. Positive numbers correspond to a uniaxial MAE. 
As one can see, a similar tendency is shown for the anisotropy in all considered rows. Starting from the lowest values, corresponding to high but planar anisotropy, for the dopants from the Group IIIA (Ga and In) the MAE moves towards positive values, with an increase of the number of valence electrons in the system. The maximal value of MAE, which is positive for As and Sb dopants, is obtained for Group VA elements. This tendency is shown for all the considered rows of elements, notifying that Group VA dopants seem to be the most promising for the consideration, specifically, As and Sb. Two other elements of this group, P and Bi, do not flip anisotropy to uniaxial (see Fig. \ref{fig:MAE_tab}). Anisotropy of Fe$_{1.5}$Mn$_{1.5}$Sn$_{0.75}$Sb$_{0.25}$ is shown with the horizontal dashed line. The number of valence electrons in the unit cell of this system is 26.75, which is out of the scale of Fig. \ref{fig:MAE_tab}. %Taking the above described tendency into account together with the result of the Mn addition we do not expect that alloying with other elements, not considered in this search around Sn in the Periodic Table, can substantially change this picture and bring new candidates to flip the easy magnetization axis to uniaxial.
The above described tendency together with the result of the Mn addition indicates that alloying with other elements around Sn in the Periodic Table, not considered in this search, can hardly bring new candidates able to flip the easy magnetization axis to uniaxial. However, more studies can be of interest.

\begin{figure} [tbp] 
%\begin{center}  
\includegraphics[scale=0.15]{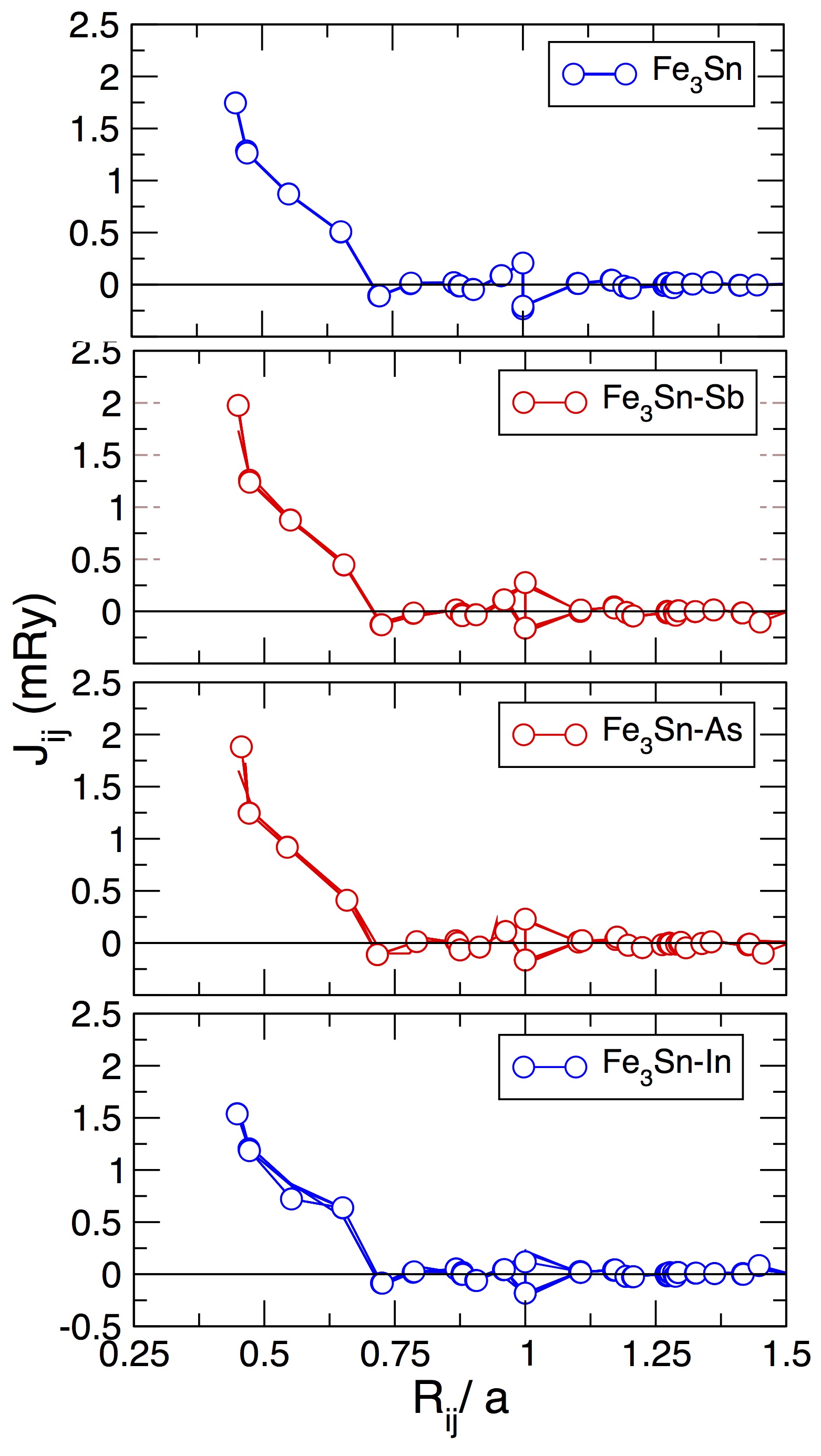} 
%\end{center} 
\caption{(Color online) Inter-site exchange parameters J$_{ij}$ between the iron atoms $i$ and $j$ separated by the distance R$_{ij}$ of pristine Fe$_3$Sn compound (a), as well as of Fe$_3$Sn$_{0.75}$M$_{0.25}$ (M=Sb (b), As (c) and In (d)). Systems with planar MAE are shown with the blue lines, while systems with uniaxial MAE are shown with red lines. $a$ stands for the lattice constant.}
\label{fig:Jij}  
\end{figure}

The dependence of the magnetic properties on the number of valence electrons in dopants was further illustrated by the calculation of the Heisenberg exchange parameters J$_{ij}$s between the atoms (see Fig. \ref{fig:Jij}). The Fe$_3$Sn system, as well as the Fe$_3$Sn$_{0.75}$M$_{0.25}$ alloys, where M=Sb, As and In, were shown to be strongly ferromagnetic as indicated by large and positive J$_0$ (defined as a sum of all $J_{ij}$'s). In Fig. \ref{fig:Jij} the systems with planar anisotropy, the pristine Fe$_3$Sn (a) and Fe$_3$Sn$_{0.75}$In$_{0.25}$ (d), which has the largest planar anisotropy, are shown in comparison with the systems with significant uniaxial MAE, namely, Fe$_3$Sn$_{0.75}$Sb$_{0.25}$ (b) and Fe$_3$Sn$_{0.75}$As$_{0.25}$ (c). 

\begin{figure} [tbp] 
%\begin{center}  
\includegraphics[scale=0.13]{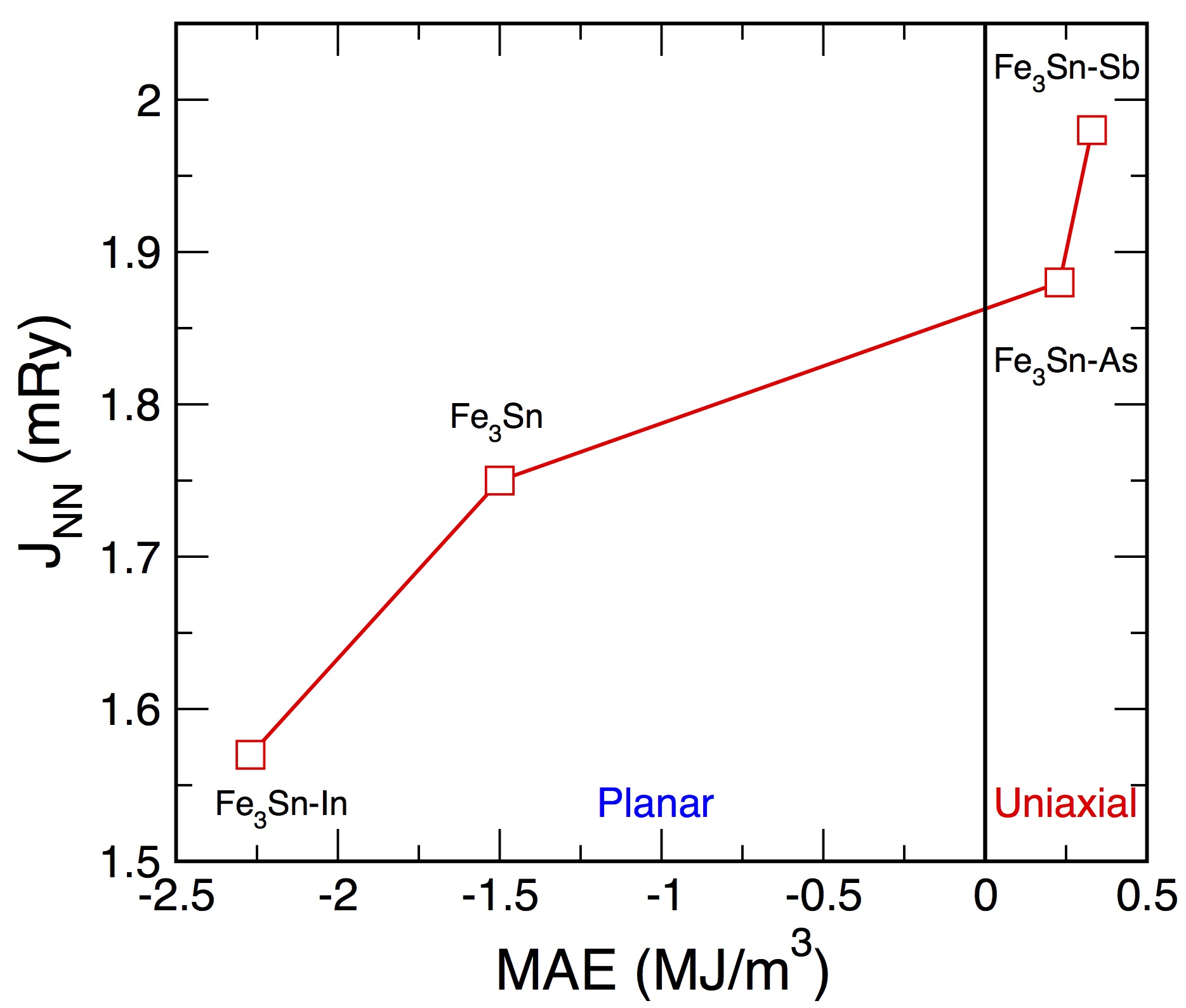} 
%\end{center} 
\caption{(Color online) Dependence of the nearest neighbor exchange parameter on the magnetocrystalline anisotropy of Fe$_3$Sn and Fe$_3$Sn$_{0.75}$M$_{0.25}$ (M=Sb, As and In). The area below zero corresponds to the planar anisotropy while the area above zero represents the uniaxial anisotropy. }
\label{fig:J_K}  
\end{figure}

Notice that the value of the nearest neighbor interaction, corresponding to the shortest inter-atomic distance, differs slightly among the considered systems. The largest nearest-neighbor first exchange interaction value was obtained for the Fe$_3$Sn$_{0.75}$Sb$_{0.25}$ system ($\sim$ 1.98 mRy) with the highest positive value of MAE. A plot of the nearest neighbor exchange interaction as function of the MAE is shown in Fig. \ref{fig:J_K}. The second highest nearest-neighbor exchange was obtained in Fe$_3$Sn$_{0.75}$As$_{0.25}$ ($\sim$ 1.88 mRy). The J$_{NN}$ value for the system with largest negative anisotropy, Fe$_3$Sn$_{0.75}$In$_{0.25}$ is the smallest of all ($\sim$1.57 mRy). The  J$_{NN}$ value for pristine binary system Fe$_3$Sn has a value of $\sim$ 1.75 mRy, which is higher than that of the system with In, but lower than the one out the systems with uniaxial anisotropy. This clearly indicates that in the presently investigated alloys, there is a certain correlation between the strength of the nearest neighbor couplings J$_{ij}$s and MAE of the system.

%) with J$_0$  $\sim$10 mRy/atom. \color{red}Accordingly, the mean-field T$_C$ is  $\sim$1050 K. \color{black}

\begin{figure} [tbp] 
%\begin{center}  
\includegraphics[scale=0.141]{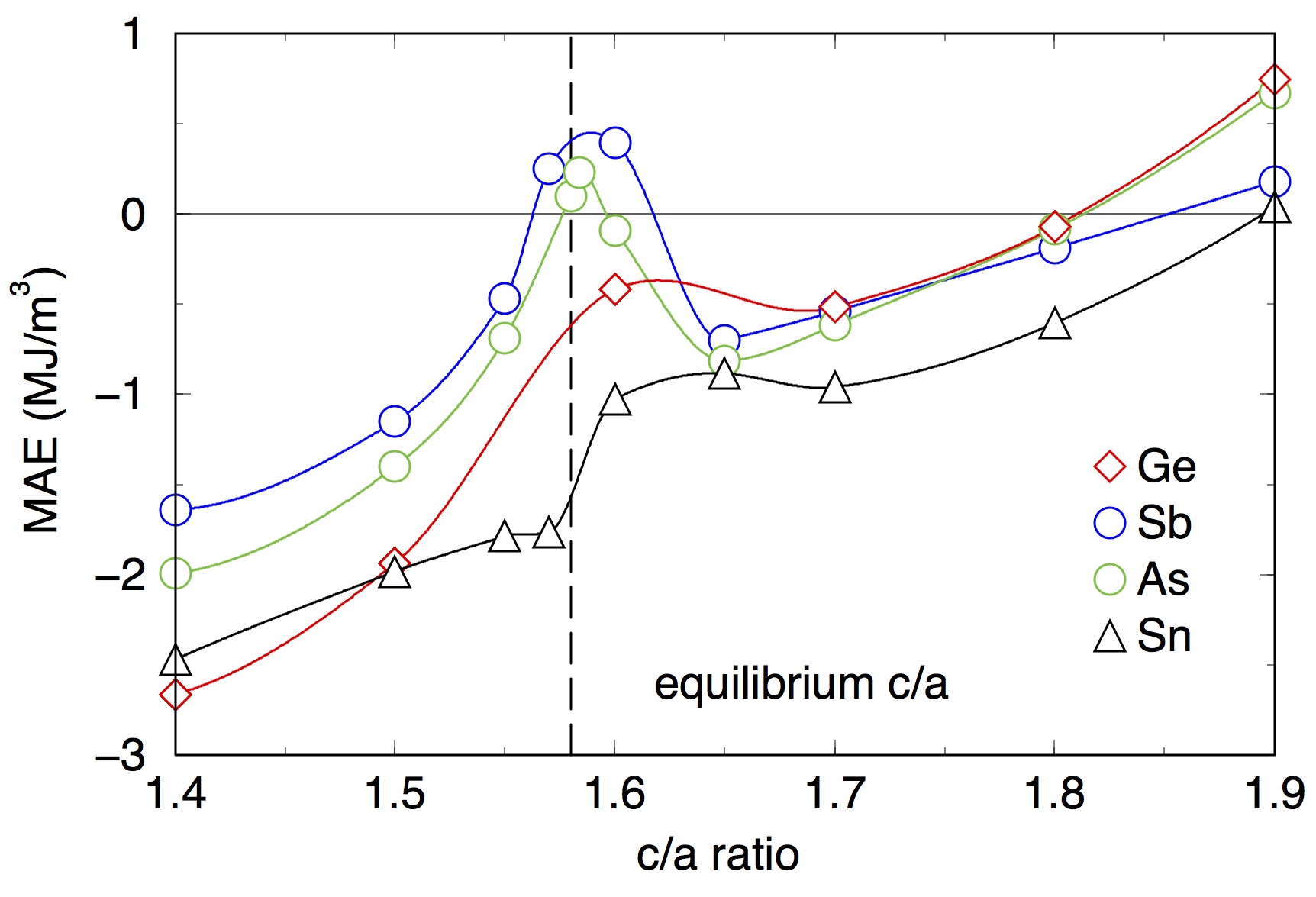} 
%\end{center} 
\caption{(Color online) Dependence of the magnetocrystalline anisotropy on the hexagonal c/a ratio for M=Sn, Ge, As and Sb impurities. The averaged equilibrium c/a ratio about 1.58 is shown with the dashed line.}
\label{fig:ca}  
\end{figure}

Taking into account that atomic relaxations substantially influence the MAE, we also studied the effect of the hexagonal lattice ratio, c/a, on the value of MAE. Fig. \ref{fig:ca} shows the  change of the MAE when the c/a ratio increases from 1.4 to 1.9 for Fe$_3$Sn$_{0.75}$M$_{0.25}$ (M= Sn, Ge, As and Sb) systems at a fixed equilibrium volume. The averaged equilibrium c/a ratio ($\sim$1.58)  is shown with the vertical dashed line. As the dopant concentration is low (6.25 at \%) the deviations from 1.58 of the c/a ratio for different dopants are rather minor. 

For all considered dopants the MAE behavior with the c/a increase is basically linear, illustrating that when the crystal is stretched, the easy axis can be switched from planar to uniaxial for all the structures. For instance, for Ge-doped Fe$_3$Sn a change of the c/a ratio at the fixed volume leads to the change of the anisotropy value from rather high with planar easy magnetization direction, $\sim$ -2.7 MJ/m$^3$, to almost  $\sim$0.7 MJ/m$^3$ with uniaxial easy magnetization direction. This strain induced change is largest for Ge doping, out of all considered dopants. 
It is worth noting that changing MAE to uniaxial requires a variation of c/a from $\sim$1.58 to 1.8, corresponding to 14 \%. In the experiment it is hardly possible to change c/a of such systems by more than few percents without structural changes, i.e. without altering the hexagonal structure. 

It is important to mention that for all the structures there is a peak of the MAE vs. c/a curve (Fig. \ref{fig:ca}) in the region near the equilibrium c/a.
 In some cases, like for M=Sb and As it can lead to the switching of the easy magnetization direction, however, this region is rather narrow and a small change in the c/a ratio (like 0.5\% from the equilibrium) can switch the axis from planar to uniaxial and back. The highest peak is for Sb- doped Fe$_3$Sn; it corresponds to the highest anisotropy value and uniaxial easy magnetization direction. 
 %As the phase of Fe$_3$Sn$_{0.75}$As$_{0.25}$ was experimentally stabilized when doped with Mn, we estimated the influence of Mn on the c/a ratio. The calculated c/a of Fe$_{0.5}$Mn$_{0.5}$Sn$_{0.75}$Sb$_{0.25}$ is equal to 1.55, while the c/a of one of the most promising (from the point of view of MAE) but unstable structures FeSn$_{0.75}$Sb$_{0.25}$ is equal to 1.57 (see Fig. \ref{fig:ca}). As one can see the value of MAE at the equilibrium c/a is equal to 0.33 MJ/m$^3$, while when c/a changes to 1.55, the value of MAE is almost zero. From this it can be predicted that the addition of Mn should not lead to the uniaxial anisotropy. This result was confirmed experimentally (\color{red}Cristina and Bahar, I need your confirmation here\color{black})

 For Ge-doped structure, as well as for the binary Fe$_3$Sn slight modifications of the  c/a near the equilibrium are not resulting in the MAE peak sufficiently high to switch the easy magnetization direction. 
 %For Ge-doped structure, as well as for the binary Fe$_3$Sn peak near the equilibrium c/a is not high enough to switch the easy magnetization direction.
 It is interesting to notice, that for these two structures the peak is not exactly at the equilibrium c/a ratio but shifted, meaning that the small change of the c/a ratio due to, for instance, alloying or heating, can  slightly increase anisotropy for some structures. 
 Therefore we underline that the MAE of Fe$_3$Sn with any of the dopants turns out to be very sensitive to the change of c/a.
 
\subsection{\label{sec:exp_res}Reactive Crucible Melting technique} %\color{red}(Bahar)}
%Bahar -beginning
The effect of substituting Sn by its neighboring elements was experimentally studied by high-throughput RCM method and the production of Fe-Sn$_x$M$_{1-x}$ crucibles, M= Sb, Si, Ga, Ge, Pb, In, Bi and $0.5<x<0.75$. The diffusion zone of the crucibles annealed at the selected temperatures in the temperature interval from 1013~K to 1073~K %[740\degree C-800\degree C] 
were carefully screened by the EDX analysis. None of the Fe$_3$(Sn,M) structures was stable and therefore not formed in any of the crucibles (for details on the results  see Supplemental Material).

Our theoretical calculations predict the change of anisotropy from planar to uniaxial in Fe$_3$Sn system by partial substitution of Sn by Sb. However, the formation enthalpy plot shows that Fe-Sn$_x$Sb$_{1-x}$ is unstable in the addressed concentration range (see Fig. \ref{fig:stab}). Besides that, it was experimentally found  that Sb destabilized the formation of 3:1 structure and instead the Fe$_3$(Sn,Sb)$_2$ phase was formed. This phase is not desirable for permanent magnet applications due to the small concentration of Fe and therefore low magnetization as well as its negligible MAE \cite{Sales_Fe3Sn, Bahar_10}. For stabilization of Fe$_3$(Sn,Sb) phase, substitution of Fe by Mn (with one less electron) was suggested to be effective to compensate an extra electron of Sb in comparison to Sn. Accordingly, the Fe-Mn$_{0.75}$Sn$_{0.5}$Sb$_{0.5}$ and Fe-Mn$_{1.5}$Sn$_{0.75}$Sb$_{0.25}$ crucibles were synthesized and annealed in the temperature range of 1013~K to 1073~K.%740\degree C to 800\degree C. 

\begin{figure}  [tbp] 
\includegraphics[scale=0.22]{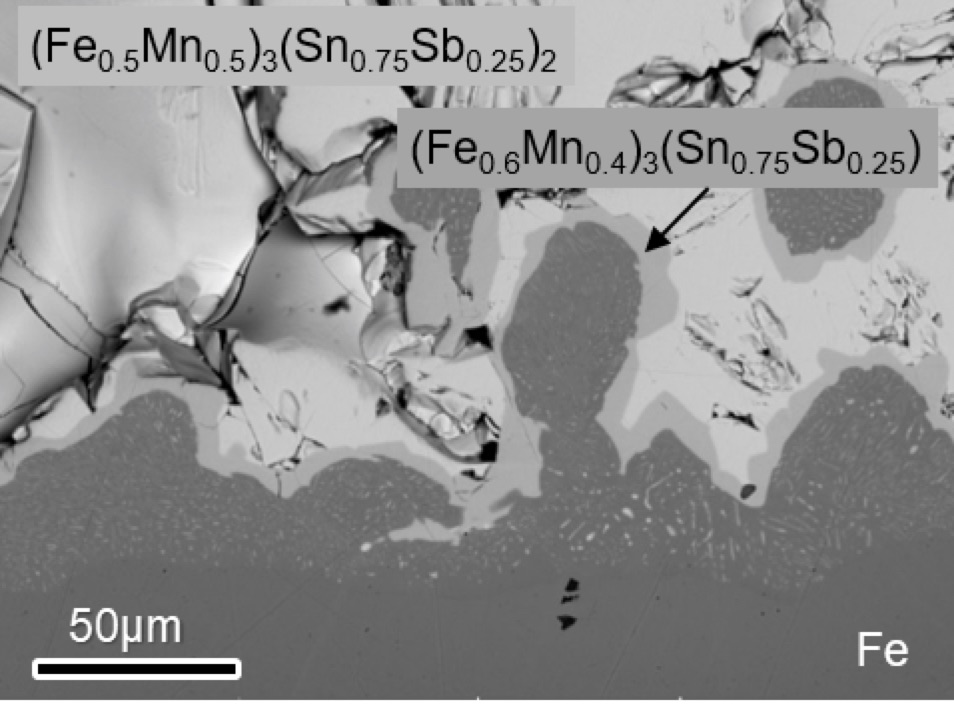} 
%\end{center} 
\caption{BSE image of the Fe- Mn$_{0.75}$Sn$_{0.75}$Sb$_{0.25}$ crucible annealed at 1013~K%740\degree C
. Narrow layer of (Fe$_{0.6}$Mn$_{0.4}$)$_3$(Sn$_{0.75}$Sb$_{0.25}$) is formed. However 3:2 phase is the dominant forming phase in the crucible. }
\label{fig:Bh}  
\end{figure}

It was found that Mn and Sb substitution preserved the hexagonal Ni$_3$Sn structure of the parent Fe$_3$Sn phase and (Fe,Mn)$_3$(Sn,Sb) phase formed in the quaternary reactive crucibles. Fig. \ref{fig:Bh} shows the microstructure forming in the diffusion zone of the Fe-Mn$_{0.75}$Sn$_{0.75}$Sb$_{0.25}$ reactive crucible annealed at 1073~K %800\degree C 
for 5 days and subsequently quenched. A thin layer of (Fe$_{0.6}$Mn$_{0.4}$)$_3$(Sn$_{0.75}$Sb$_{0.25}$) phase is formed and bordered by Fe crucible. In addition, (Fe$_{0.5}$Mn$_{0.5}$)$_3$(Sn$_{0.75}$Sb$_{0.25}$)$_2$ phase is formed in large quantity on top of the 3:1 phase. The phase stability of the Sb-Mn doped Fe$_3$Sn compound is confirmed by RCM. However, to measure magnetic properties, such as M$_s$, T$_\mathrm{C}$, and MAE; a single-phase material has to be synthesized. That was done by the solid state reaction (SSR).
%[\color{red}Bahar and Cristina, we need some motivation why do we need two experiments\color{black}]
%… I may add some sentences after I get the whole paper
%…We should see if it makes sense to state that all these substitutions were done experimentally as Uppsala only found Sb effective (uniaxial).  Supplemental Material will be written (if needed)
%… I made a new RCM crucible to get larger diffusion zone to be able to observe the domain structure (need time)
%… Therefore, the image attached will be changed

%Bahar ending

\subsection{\label{sec:exp_res_c}Solid State Reaction} %  \color{red}(Cristina)}

\begin{figure}  [tbp] 
\includegraphics[scale=0.11]{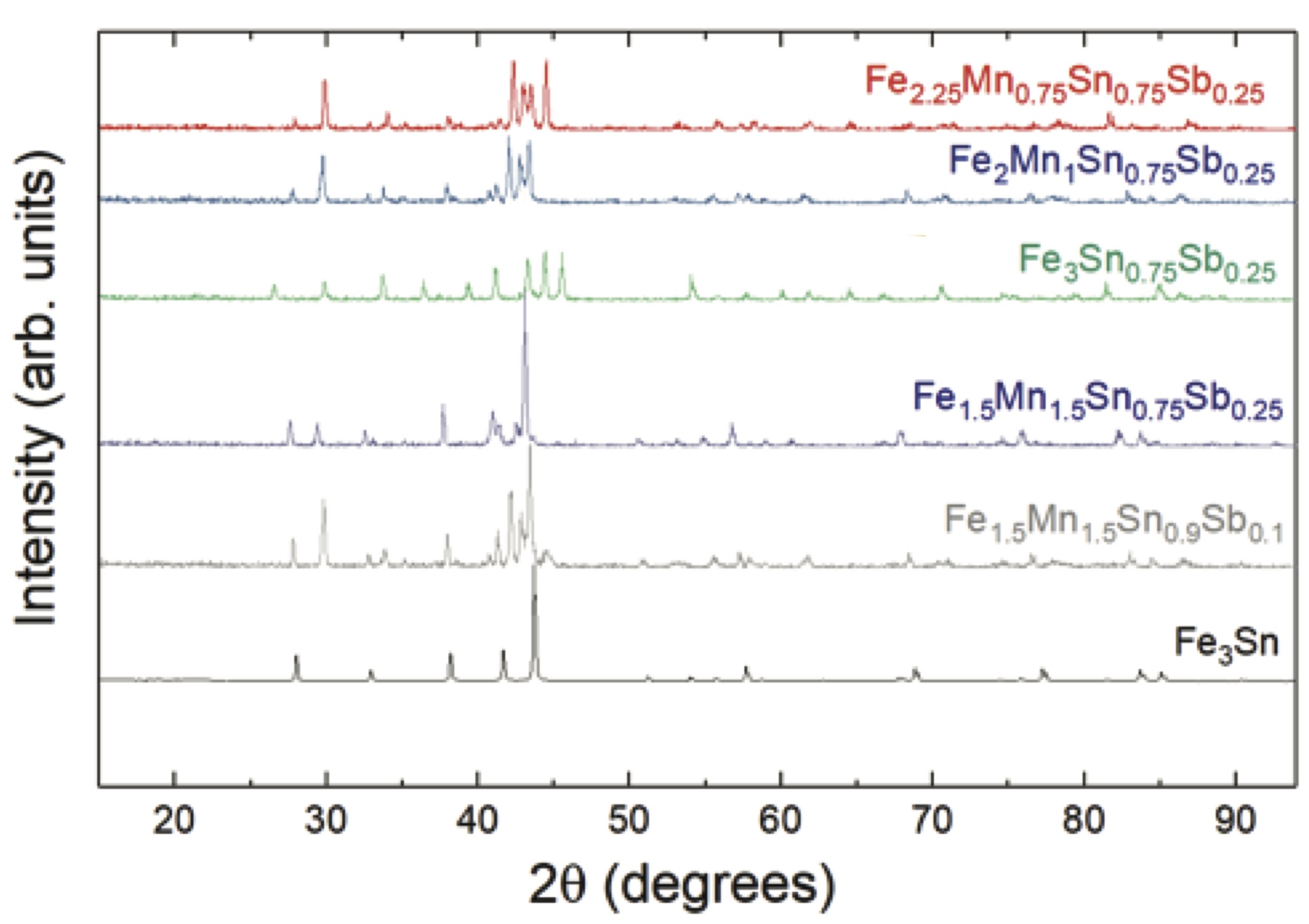} 
%\end{center} 
\caption{(Color online) XRD patterns of the Fe$_{y}$Mn$_{3-y}$Sn$_{x}$Sb$_{1-x}$ samples.}
\label{fig:Cr1}  
\end{figure}

Several Fe$_{y}$Mn$_{3-y}$Sn$_{x}$Sb$_{1-x}$ samples (y=3 and x=1; y= 2.25, 2, 1.5 and x=0.75; and y=1.5 and x = 0.9) with different concentrations of Mn and Sb were prepared by solid state reaction. %Structural characterization
XRD patterns of the produced samples by two subsequent SSRs are shown in Fig. \ref{fig:Cr1}. For all the samples, the 3:1 phase was found though only for the parent Fe$_3$Sn alloy we found a single phase. %This could be due to the fact that we did not use the proper synthesis parameters during the SSR \color{red}[Cristina, we need to discuss that. Why the proper parameters were not used?]\color{black}. 
Rietveld refinements were performed on all the XRD patterns and the results are shown in Table \ref{tab:lat}.

%\begin{figure}  [tbp] 
%\includegraphics[scale=1.2]{Cristina_2.pdf} 
%\end{center} 
%\caption{(color online). Lattice parameters of the Fe$_{3-x}$Mn$_{x}$Sn$_{1-y}$Sb$_{y}$ alloys, obtained by Rietveld refinements of the XRD patterns }
%\label{fig:Cr2}  
%\end{figure}

\begin{figure}  [tbp] 
\includegraphics[scale=0.1]{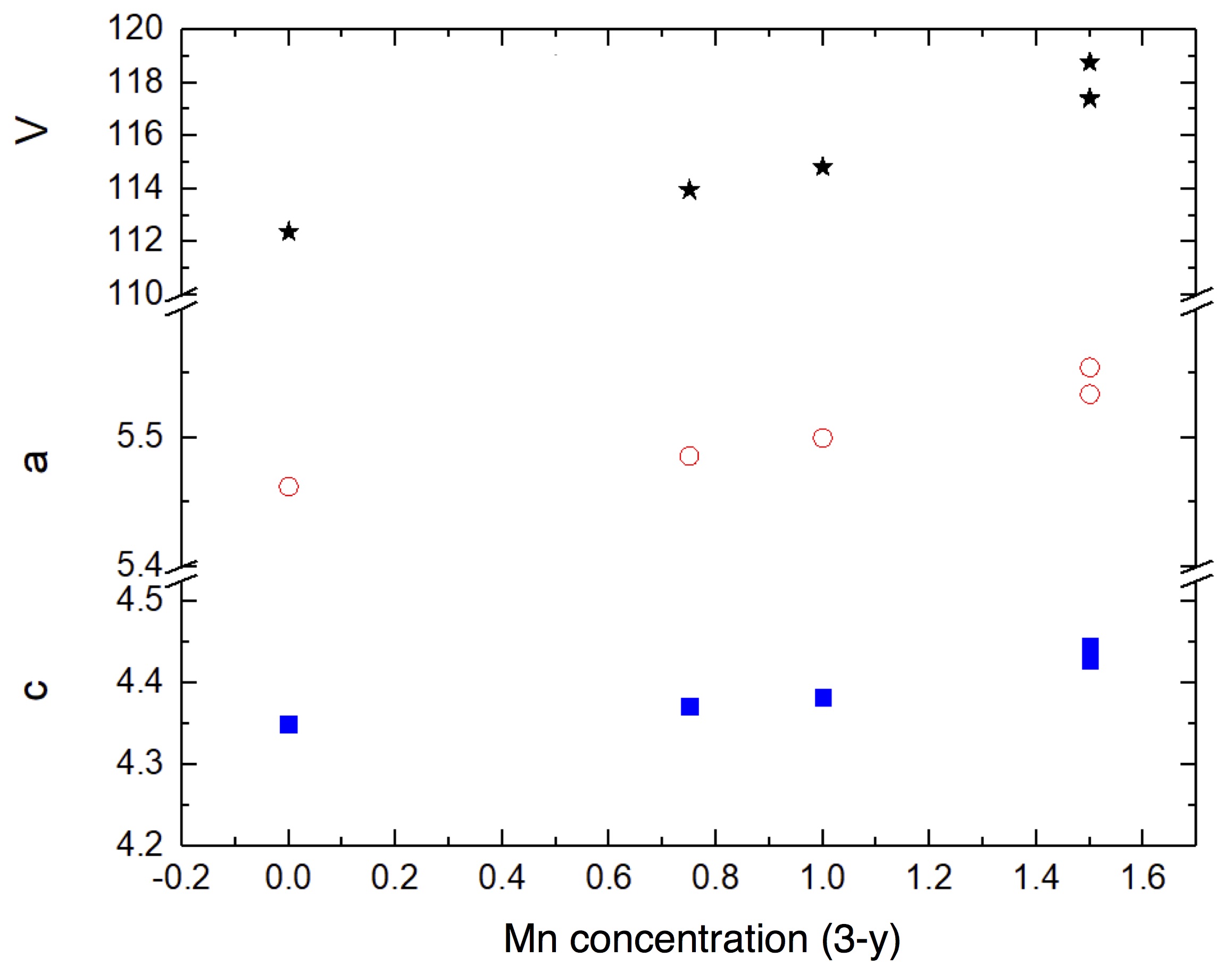} 
%\end{center} 
\caption{(Color online) Evolution of the lattice parameters and volume, obtained by Rietveld refinements of the XRD patterns, of the Fe$_{y}$Mn$_{3-y}$Sn$_{x}$Sb$_{1-x}$ alloys. These values are also shown in Table 1. }
\label{fig:Cr3}  
\end{figure}

\begin{table}[htbp]
\begin{center}
  \begin{tabular}{c c r c c}
    \hline \hline      
     \rule{0pt}{3ex}Sample & a($\mathrm{\mathring{A}}$) & c($\mathrm{\mathring{A}}$) \\
    \hline
    \\
    Fe$_{3}$Sn & 5.4621(5) & 4.3490(6) \\ 
     Fe$_{2.25}$Mn$_{0.75}$Sn$_{0.75}$Sb$_{0.25}$  &5.4858(5) &4.3721(6)\\ 
    Fe$_{2}$Mn$_{1}$Sn$_{0.75}$Sb$_{0.25}$   &5.5000(4) &4.3829(6) \\
    Fe$_{1.5}$Mn$_{1.5}$Sn$_{0.75}$Sb$_{0.25}$   &5.5338(1) &4.4270(2)\\
    Fe$_{1.5}$Mn$_{1.5}$Sn$_{0.9}$Sb$_{0.1}$   &5.5545(3) &4.4453(4)\\
     \hline \hline
  \end{tabular}
\end{center}
\caption{Lattice parameters of the Fe$_{y}$Mn$_{3-y}$Sn$_{x}$Sb$_{1-x}$ alloys, obtained by Rietveld refinements of the XRD patterns}
\label{tab:lat} 
\end{table}
The lattice parameters of the alloys (see Table \ref{tab:lat} and Figure \ref{fig:Cr3}) follow an increasing trend for higher values of the Mn concentration. %However, as one can see from Fig. \ref{fig:Cr3} for a certain content of Mn (3-y=0.5) the values of the lattice parameters are far from the observed trend. Actually, these are the highest out of all the series. We did not find any explanation for this.\color{red} [Cristina, we need to discuss that]\color{black}.  
The dependence of the lattice parameters on the concentration of Sb was not studied as only two points were available.

%Magnetic characterization

\begin{figure}  [tbp] 
\includegraphics[scale=0.12]{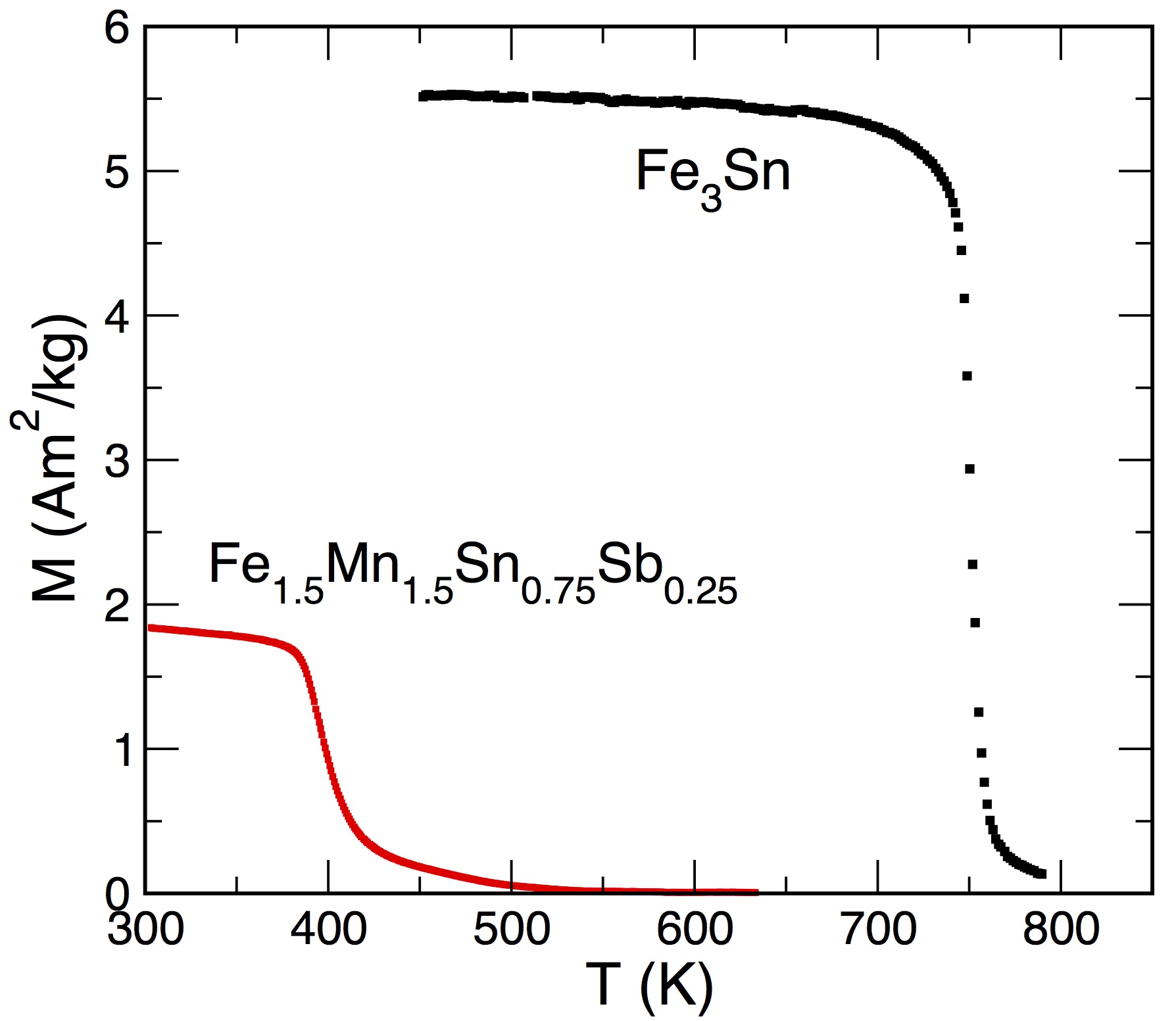} 
%\end{center} 
\caption{(Color online) M(T) curves for selected compositions within the Fe$_{y}$Mn$_{3-y}$Sn$_{x}$Sb$_{1-x}$ alloys. The abrupt drop corresponds to the Curie temperature, determined precisely by the derivative of the curve.}
\label{fig:Cr4}  
\end{figure}

Temperature dependent magnetization curves M(T) were measured for two selected samples with the highest amount of the 3:1 phase, namely, Fe$_3$Sn, and with equal concentration of Fe and Mn, Fe$_{1.5}$Mn$_{1.5}$S$_{0.75}$Sb$_{0.25}$. The M(T) curves are shown in Fig. \ref{fig:Cr4}. The calculation of the Curie temperatures were performed by the derivative of the M(T). 
The obtained Curie temperature of  Fe$_3$Sn system, T$_C$=748~K,  is rather high and in very good agreement with the existing experimental data (T$_C$=743~K \cite{T_C_exp} and  T$_C$=725~K \cite{Sales_Fe3Sn}), while that of the diluted sample, 393~K, is much lower than that of the parent composition Fe$_3$Sn. Similar values were found previously in literature for Fe$_{1.5}$Mn$_{1.5}$Sn$_{0.9}$Sb$_{0.1}$ and Fe$_{1.5}$Mn$_{1.5}$Sn$_{0.85}$Sb$_{0.15}$, with T$_C$ = 405~K %132\degree C
 \cite{Sales_Fe3Sn}. %However, for the other compound with small Mn concentration, Fe$_{2.5}$Mn$_{0.5}$Sn$_{0.75}$Sb$_{0.25}$, the T$_C$ obtained in this work \color{red}is substantially  lower [How do we explain that?] \color{black}than that of similar compounds, found in Ref. \cite{Sales_Fe3Sn}, namely,  T$_C$ = 645~K %372\degree C 
 %for Fe$_{2.5}$Mn$_{0.5}$Sn$_{0.95}$Sb$_{0.05}$.

In order to examine how the addition of Mn to the Fe$_3$(SnSb) compound affects the anisotropy, the domain structure of the formed 3:1 phase was investigated. For the ferromagnets with uniaxial anisotropy observation of characteristic domain patterns, namely stripe or branched domains, is expected. However, our Kerr analysis on the formed Mn/Sb substituted 3:1 phase shows non-uniaxial domain structure. This observation agrees with the negative theoretical value of anisotropy for the  Fe$_{1.5}$Mn$_{1.5}$Sn$_{0.75}$Sb$_{0.25}$ system.

\subsection{\label{sec:micromagn}Micromagnetic simulations} %\color{red}(Thomas)\color{black}}

\begin{figure}
\includegraphics[scale=0.25]{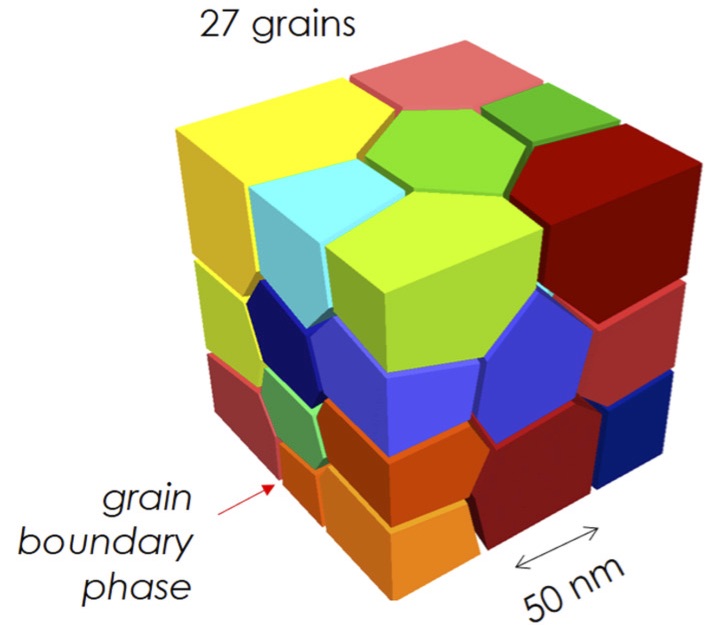}
\caption{\label{Fig_Microstructure}(Color online) Synthetic microstructure used for the simulation of the demagnetization curve of a nanocrystalline Fe$_{3}$Sn$_{0.75}$Sb$_{0.25}$ magnet.}
\end{figure}

A micromagnetic model was developed in order to estimate the energy density product $(BH)_\mathrm{max}$ and coercive field $\mu_{0}H_\mathrm{c}$ of a potential magnet made of Fe$_{3}$Sn$_{0.75}$Sb$_{0.25}$. Using the software tool Neper \cite{quey2011large} we created a synthetic microstructure based on Voronoi tessellation as shown in Figure \ref{Fig_Microstructure}. 

The granular grain structure consisted of 27 grains with an average grain diameter of 50 nm. Each grain’s spontaneous magnetization, anisotropy energy density were taken from ab-initio calculations ($\mu_{0}M_\mathrm{s,grain} = 1.52$~T, $K_\mathrm{u,grain} = 0.33$~MJ/m$^{3}$). The exchange stiffness constant was assumed to be $A_\mathrm{ex,grain} = 10$~pJ/m. The grain’s easy axes were randomized within a cone angle of 5° with respect to positive z-axis. 

Between the grains we assumed a 4 nm thick iron-rich ferromagnetic grain boundary phase ($K_\mathrm{u,gb}=0$). Its spontaneos magnetization was assumed to be M$_{s,gb}$=0.81~T. Accordingly, the exchange stiffness constant was reduced to A$_{ex,gb}$ = 3.7~pJ/m.
%Between the grains a 4 nm thick we assumed a ferromagnetic grain boundary phase ($K_\mathrm{u,gb}=0$)\color{red}[!]\color{black}. It’s spontaneous magnetization was assumed to be $\mu_{0}M_\mathrm{s,gb}=0.81$~T. Accordingly, the exchange stiffness constant was reduced to $A_\mathrm{ex,gb} = 0.37$~pJ/m. 
%\color{red} How do you estimate A? For hexagonal lattices there are at least 2 definitions. May be we need to add the equation? \color{black}

The magnetization value of the grain boundary phase was obtained by numerical optimization. We used the numerical optimization framework Dakota \cite{adams2009dakota} and maximized the energy density product using $\mu_{0}M_\mathrm{s,gb}$ as a free parameter. The volume fraction of the grain boundary phase was 10 percent. The model was discretized into a uniform mesh with an edge length of 2 nm. A finite element energy minimization code \cite{fischbacher2017nonlinear} was used to compute the static hysteresis properties of the proposed model. 

In order to compute the demagnetization curve we calculated the equilibrium states for a subsequently decreasing external field $H_\mathrm{ext}$. The field step $\mu_{0}\Delta H_\mathrm{ext} = -1$~mT. In order to compute the $B(H)$ loop and the expected energy density product we corrected the loop which was obtained for a magnet with cubic shape with the macroscopic demagnetization factor $N = 1/3$. 

\begin{figure}
\includegraphics[scale=0.3]{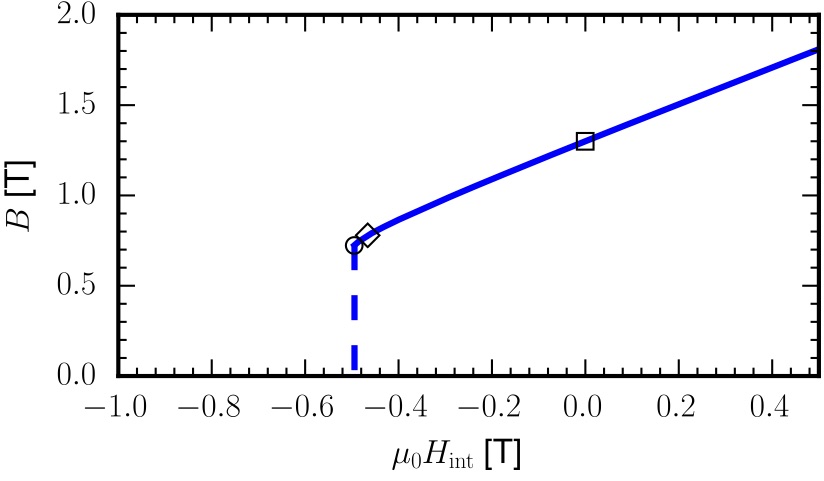}
\caption{\label{fig:Tom_3}(Color online) Computed $B(H)$ curve (magnetic induction as function of the internal field) for Fe$_{3}$Sn$_{0.75}$Sb$_{0.25}$. Remanence, energy density product, and coercive field are marked with a rectangle, diamond, and  circle respectively.}
\end{figure}

\begin{figure}
\includegraphics[scale=0.12]{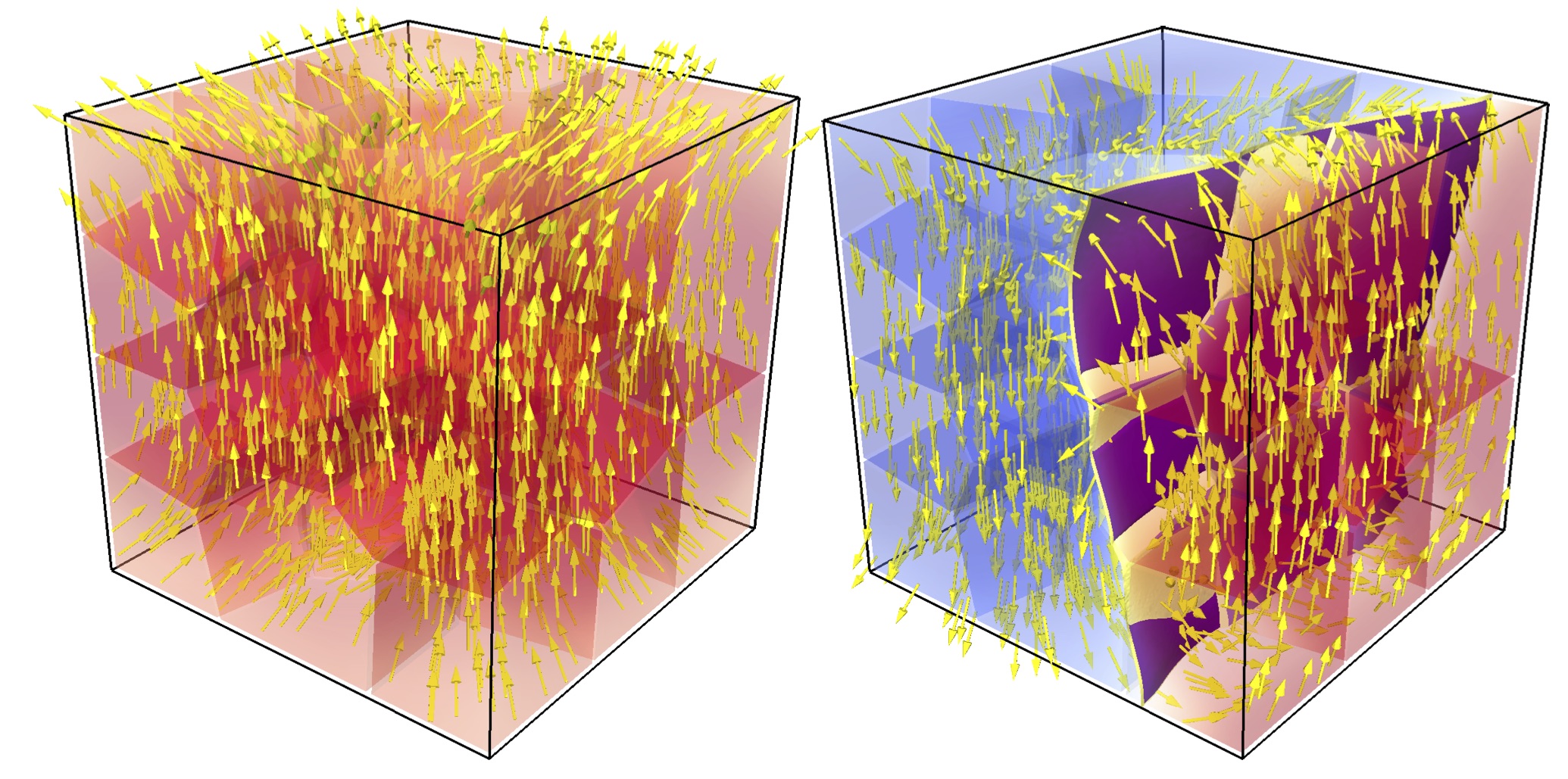}
\caption{\label{fig:Tom_2}(Color online) During magnetization reversal domain walls become pinned near the grain boundaries.}
\end{figure}

Figure \ref{fig:Tom_3} shows the computed $B(H)$ curve and Fig. \ref{fig:Tom_2} illustrates the equilibrium magnetic states before and after the first switching event, respectively.  The computed coercive field is $\mu_{0}H_\mathrm{c} = -0.49$~T and the computed energy density product is 290~kJ/m$^{3}$. %\color{red} Can we scale this to a known system? Nd2Fe14B or similar? \color{black} 
This is about 3/4 of the maximum energy density product reported for commercially available  Nd-Fe-B magnets \cite{Oliver, alex_2p, alex_1p}.

%That is close to the energy density product of the best known magnet Nd-Fe-B, estimated to be higher than 200~kJ/m$^{3}$ \cite{alex_1p} (with the maximum energy density product, above 400~kJ/m$^{3}$, reported for Nd$_2$Fe$_{14}$B  \cite{Oliver, alex_2p, alex_1p}).  

%\begin{figure}
%\includegraphics{Fig_Reversal.eps}
%\caption{\label{Fig_Reversal} Left: Computed $B(H)$ curve (magnetic induction as function of the internal field) for Fe$_{3}$Sn$_{0.75}$Sb$_{0.25}$. Right: During magnetization reversal domain walls become pinned near the grain boundaries.}
%\end{figure}  

%\begin{figure}
%\centering
%\begin{subfigure}{.5\textwidth}
%  \centering
%  \includegraphics[width=.2\linewidth]{Thomas_3.pdf}
%  \caption{A subfigure}
%  \label{fig:sub1}
%\end{subfigure}%
%\begin{subfigure}{.5\textwidth}
%  \centering
%  \includegraphics[width=.2\linewidth]{Thomas_3.pdf}
%  \caption{B subfigure}
%  \label{fig:sub2}
%\end{subfigure}
%\caption{A figure with two subfigures}
%\label{fig:test}
%\end{figure}

\section{\label{sec:conclusions}Discussion and Conclusions}

Electronic structure and magnetic properties of the hexagonal Fe$_{3}$Sn compound doped with 6.25 at \% of Si, P, Ga, Ge, As, Se, In, Sb, Te, and Bi were studied theoretically from first principles and experimentally. 
Our calculations show that at low concentrations of some dopants, such as Ga and Ge, the considered phases are stable. In contrast to that, in the case of the Sb dopant the phase was shown to be unstable against decomposition into the mixture of pure elements. Our experimental study using RCM method supports this theoretical prediction that the addition of Sb into Fe-Sn system destabilizes the formation 3:1, however further doping of Mn into the Fe sublattice stabilizes the structure. The SSR technique confirmed stabilization of the 3:1 phases with different concentrations of Mn.

Theoretical simulations predict that doping with As, Sb and Te can change the easy magnetization direction to uniaxial. However, the change of the c/a ratio also substantially influences the MAE. In the case of Sb and As the peak in the MAE vs c/a dependence leads to the change of the easy magnetization axis from planar to uniaxial in the region close to the equilibrium c/a. For the Sb and As dopants, the region where anisotropy is uniaxial, is rather narrow and even a small change of c/a, for instance due to alloying with small amounts of Mn, can lead to the switch of the easy magnetization direction.

As follows from our calculations, the change of anisotropy due to addition of Mn to Fe$_{3}$Sn$_{0.75}$Sb$_{0.25}$ system turns uniaxial anisotropy back to planar. This estimation indicates that the predicted uniaxial anisotropy can hardly be observed experimentally. On one side the presence of Mn stabilizes the Sb-doped alloy, but on the other side it basically flips the anisotropy back to the planar value of an undoped system. The experimentally stabilized (FeMn)$_3$SnSb phase shows non-uniaxial domain structure in nice agreement with the theoretical prediction.%The experimentally stabilized (FeMn)$_3$SnSb phase was found to have planar anisotropy in nice agreement with the theoretical prediction.

We studied all the dopants around Sn in the Periodic Table that can occupy the Sn sublattice and can likely be mixed with Fe$_3$Sn. From the MAE data grouped in the way as these dopants are placed in the Periodic Table of elements one can see the tendency to increase the value of MAE from the III to the V group with the increase of the number of valence electrons. However, in the VI group there is a slight decrease for Te and Se, respectively. Thus the Group V seems to be the most promising out of all the considered groups and the Sb addition gives the largest uniaxial anisotropy. Looking at the vertical distribution in the Periodic Table, it appears that MAE increases from P to As and to Sb, but then for Bi there is a clear drop. We conclude that the most preferable choice of dopants should be the column with As and Sb, and we do not expect that other dopants can substantially improve the desired properties of the Fe$_3$Sn compound.

Further, the micromagnetic simulations allowed us to estimate the magnetic induction of the most promising system, Fe$_{3}$Sn$_{0.75}$Sb$_{0.25}$, as a function of the internal field. The computed coercive field is equal to -0.49 T. The calculated density energy product, 290~kJ/m$^{3}$, is at the level of best known to date magnets.

We investigated the Fe$_{3}$Sn$_{0.75}$M$_{0.25}$ system by different theoretical approaches as well as experimentally, in order to get a wider view on the magnetic properties of the system. Certain dopants, like Sb, can turn MAE of the hexagonal Fe$_{3}$Sn uniaxial. Furthermore, other magnetic properties of this system, such as saturation magnetization of 1.51~T and energy density product compatible with the values of the best known magnets are impressively high. However, such a turn  of MAE is very sensitive to lattice deformations and the value of the magnetocrystalline anisotropy is reduced compared to the one of the parent phase. Further, the hexagonal phase becomes unstable with respect to decomposition. Addition of Mn allowed us to stabilize the system with dopants experimentally, however, the anisotropy turned back to planar. Therefore further search for better stabilizers or their combinations (the so-called co-doping) might be considered in order to find the compromise between the stability and uniaxial MAE.

%In all the considered structures at c/a around 1.9, what corresponds to approximately 20\% stretching, the switch of the easy magnetization direction is observed. 
 
% From the calculations of exchange-interactions it was shown that the Fe$_3$Sn system is strongly ferromagnetic.

\section*{Acknowledgments}
Authors  acknowledge  support  from  NOVAMAG  project,  under  Grant  Agreement  No.  686056,  EU  Horizon  2020  Framework  Programme. The computations were performed on resources provided by the Swedish National Infrastructure for Computing (SNIC) at PDC and NSC centers. O.E. acknowledges support from STandUPP, eSSENCE, the Swedish Research Council and the KAW foundation (grants 2012.0031 and 2013.0020). O.G. acknowledges the Hessen LOEWE Response programme. Authors are thankful to Yaroslav Kvashnin for useful discussions.
%\bibliography{bibl_copy}

%

\newpage
\clearpage
\renewcommand{\thefigure}{S\arabic{figure}}
\renewcommand*{\bibnumfmt}[1]{[S#1]}
\renewcommand*{\citenumfont}[1]{S#1}
\setcounter{figure}{0}   
\section*{Supplemental material for ``Tuning magnetocrystalline anisotropy of $\mathrm{Fe_3Sn}$ by alloying''}

\subsection*{Theoretical calculations}
\begin{figure}[hbp]
\centering
\includegraphics[scale=0.15]{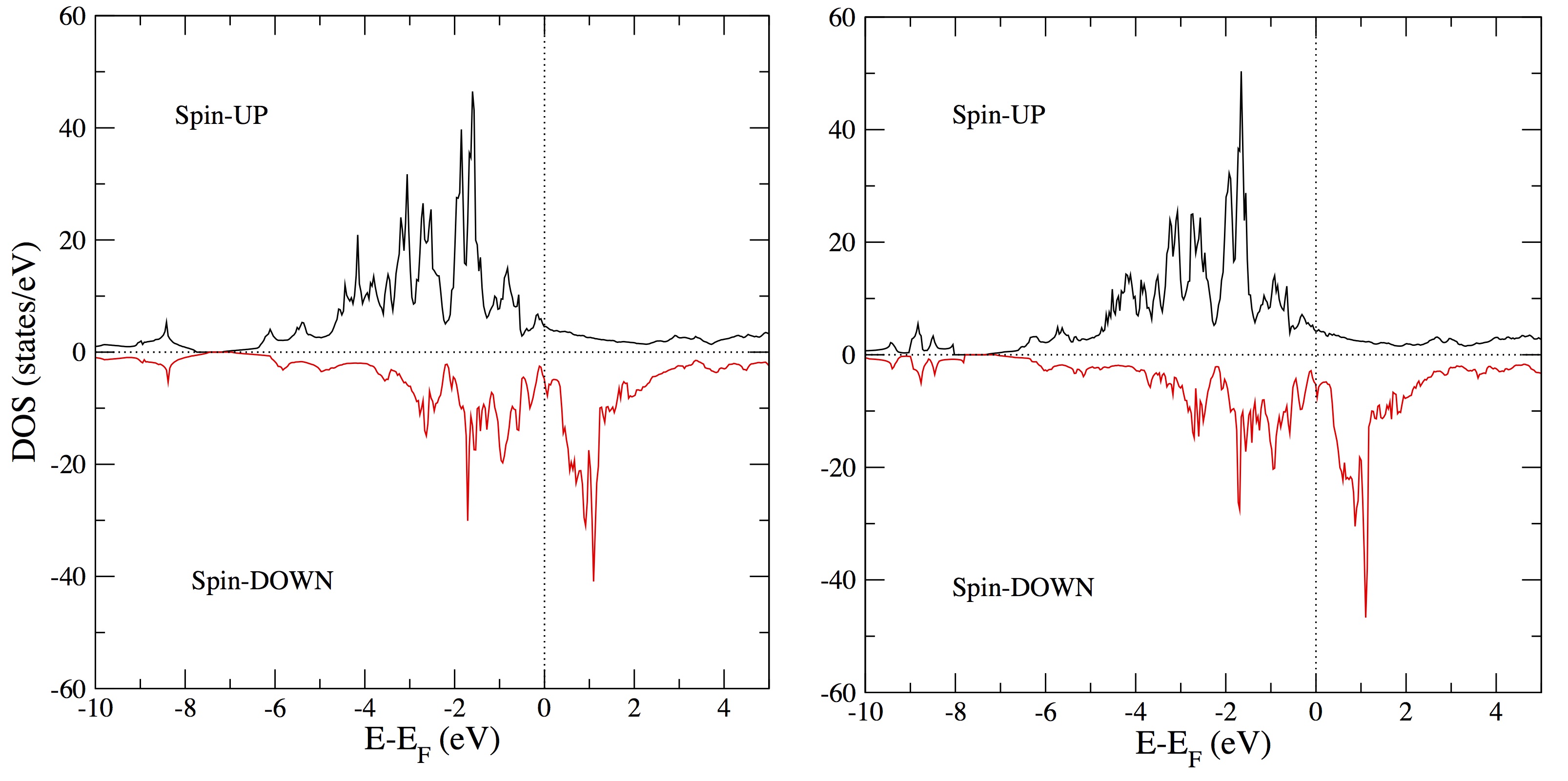}
\caption{\label{fig:DOS_1} Density of states of the Fe$_3$Sn and  Fe$_3$Sn$_{0.75}$Sb$_{0.25}$ compounds without spin-orbit coupling taken into account.}
\end{figure}

\begin{figure}[tbp]
\centering
\includegraphics[scale=0.15]{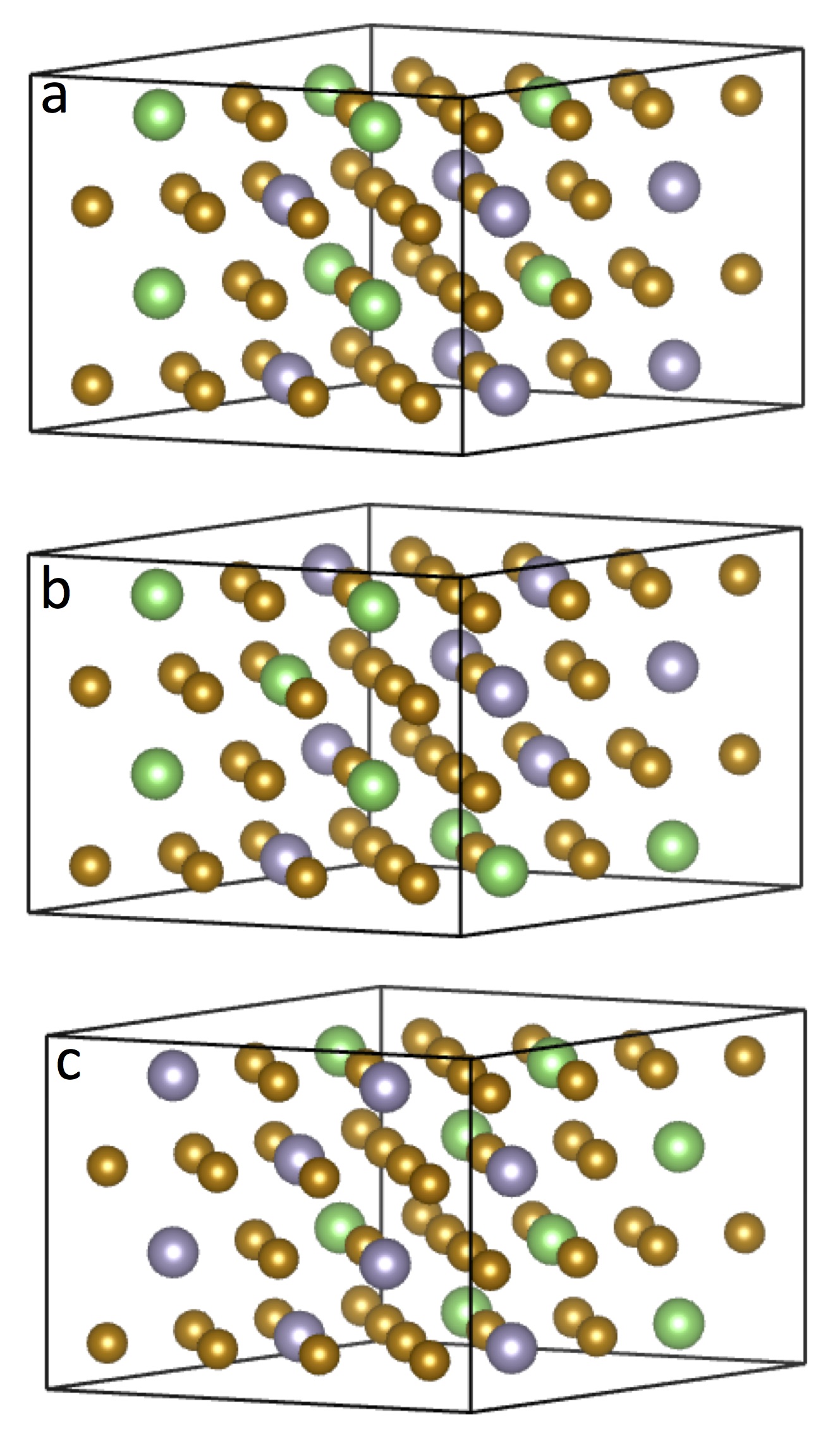}
\caption{\label{fig:strucs} Structure of the Fe$_3$Sn$_{0.75}$M$_{0.25}$ system. Green spheres (impurity atoms) are distributed on the Sn sublattice, shown with the grey spheres (host atoms) in (a) ordered, (b) random or (c) separated ways.}
\end{figure}

Figure \ref{fig:DOS_1} shows the calculated density of states for the Fe$_3$Sn and Fe$_3$Sn$_{0.75}$Sb$_{0.25}$ compounds, respectively. Figure \ref{fig:strucs} shows the three tested distributions of M impurity atoms on the Sn sublattice in the Fe$_3$Sn$_{0.75}$M$_{0.25}$ compound. For the sake of simplicity only the results for 50 at. \% of impurities is shown in all the cases. As one can see in the case of the ordered distribution, (a), impurity atoms are placed in layers mixed with the layers of the Sn atoms. In the case of the random alloy, (b) atoms are distributed in an SQS-like manner with the short-range order close to 0 for a few first coordination shells. For the case of the separated distribution of atoms, (c), the cell is split into two parts and one is fully occupied with the M atoms, while the other one is occupied exclusively with the Sn atoms. According to our estimation, for the low concentration of dopants, the ordered distribution is the most energetically preferable.

%\begin{figure}[tbp]
%\centering
%\includegraphics[scale=0.5]{DOS_fe3snsb.pdf}
%\caption{\label{fig:DOS_2} Density of states of Fe$_3$Sn$_{0.75}$Sb$_{0.25}$ system without spin-orbit coupling}
%\end{figure} 

\subsection*{RCM measurements}

\begin{figure}[tbp]
\centering
\includegraphics[scale=0.3]{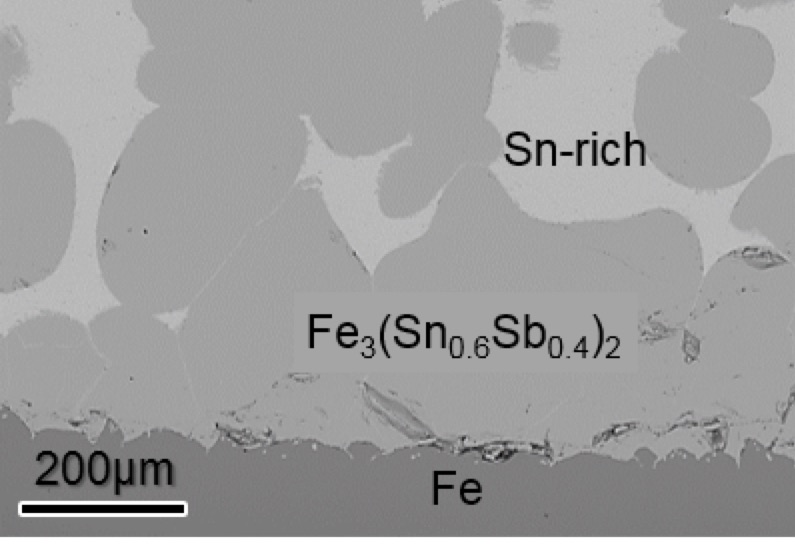}
\caption{\label{fig:bahar_s1} BSE image of the Fe-Sn$_{0.85}$Sb$_{0.25}$ crucible annealed at 1073~K. Substitution of Sb led to formation of Fe$_3$(SnSb)$_2$ and deformation of the desirable Fe$_3$Sn phase.}
\end{figure}

Using the RCM method, Fe-SnM crucibles with M= Sb, Ga, Ge, Si, In, Bi and Pb were synthesized. The crucibles were annealed in the temperature range of 1013~K to 1073~K for 5 days and subsequently quenched.
Theoretical calculations predict that doping of Sb changes the planar anisotropy in the Fe$_3$Sn compound to uniaxial. Our experimental study using RCM shows a destabilization of the 3:1 phase in Fe-Sn$_{1-x}$Sb$_x$, $0.15<x<0.5$ crucibles. The 3:1 phase did not form in any of the synthesized samples, however, instead the 3:2 phase was stabilized.  Figure \ref{fig:bahar_s1} shows the BSE image of the diffusion zone in Fe-Sn$_{0.85}$Sb$_{0.15}$ reactive crucible annealed at 1073~K. The only intermetallic compound formed in the crucible is the 3:2 phase.

\begin{figure}[tbp]
\centering
\includegraphics[scale=0.3]{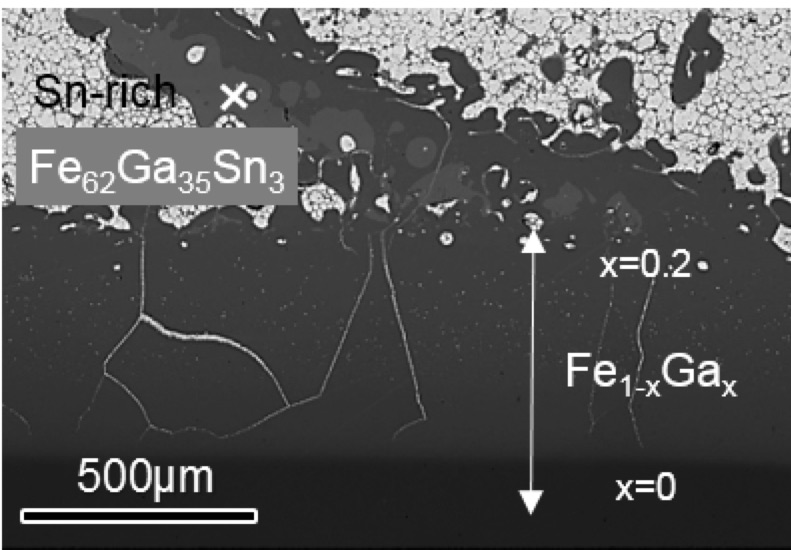}
\caption{\label{fig:bahar_s2} BSE image of the diffusion zone in the Fe-Sn$_{0.5}$Ga$_{0.5}$ crucible annealed at 1073~K. Sn remained almost unreacted and the only ternary intermetallic compound Fe$_{62}$Ga$_{35}$Sn$_3$ is formed.}
\end{figure}

\begin{figure}[tbp]
\centering
\includegraphics[scale=0.3]{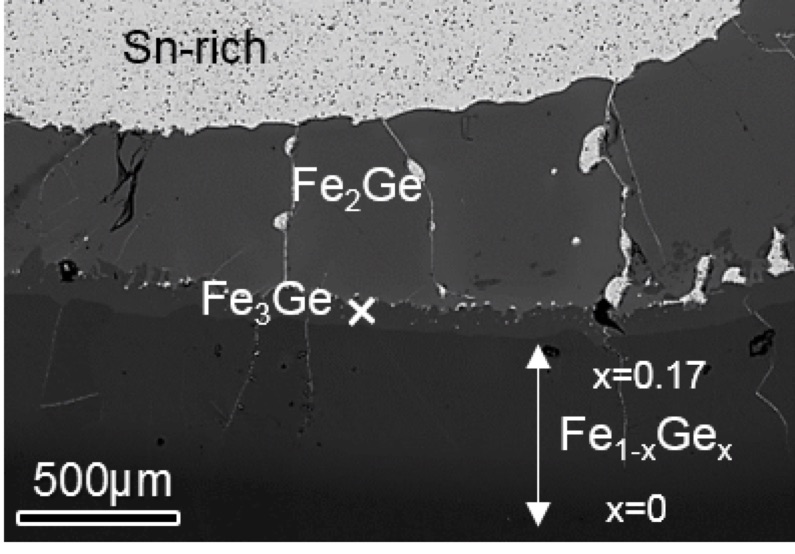}
\caption{\label{fig:bahar_s3} BSE image of the diffusion zone in the Fe-Sn$_{0.5}$Ge$_{0.5}$ crucible annealed at 1073~K. Sn remained almost unreacted and no ternary intermetallic compounds are formed.}
\end{figure}

In order to check whether substituting Ge and Ga in the Fe$_3$Sn compound preserves the parent structure, Fe-SnGe(Ga) crucibles were synthesized. The diffusion zone of the crucibles annealed at a specific temperature of 1013~K are shown in figure \ref{fig:bahar_s2} for Fe-Sn$_{0.5}$Ga$_{0.5}$ and in figure \ref{fig:bahar_s3} for Fe-Sn$_{0.5}$Ge$_{0.5}$. The observed microstructure in Fe-SnGa crucible contains a solid solution Fe$_{1-x}$Ga$_x$ region with x up to 20~at.~\%. The remaining Ga has formed the Fe$_{62}$Ga$_{35}$Sn$_3$ phase (composition found by EDX) in the diffusion zone and the Sn content is left mainly unreacted. The crucibles annealed at the higher temperatures (not shown here) contained no intermetallic compounds and only solid solution Fe$_{1-x}$Ga$_x$ was found. In the Ge containing crucible, Sn remained totally unreacted. On the Fe-rich side of the crucible, a solid solution Fe$_{1-x}$Ge$_x$ region with x up to 17 atomic percent was formed. Moreover, distinct layers of Fe$_3$Ge and Fe$_2$Ge intermetallic compounds were formed on top of the solid solution region. A very similar microstructure was observed for the crucibles annealed at higher temperatures up to 1073~K. Considering the above mentioned conditions, the stabilization of 3:1 compound was not possible.

Further experiments have been performed in order to stabilize the Fe$_3$(SnM) compound with M= Si, In, Bi, Pb and Sn:M varied from 2:1 to 1:1. Our RCM study shows that no reaction has been occurred for the 3 latter substitutions after annealing at the selected temperature range. Addition of Si to the Fe-Sn crucible mimics the microstructure forming in the Fe-Sn binary crucibles \cite{S1} whereas Si totally dissolves in Fe and does not diffuse to any of the formed Fe-Sn binary intermetallic compounds. 

\begin{figure}[tbp]
\centering
\includegraphics[scale=0.3]{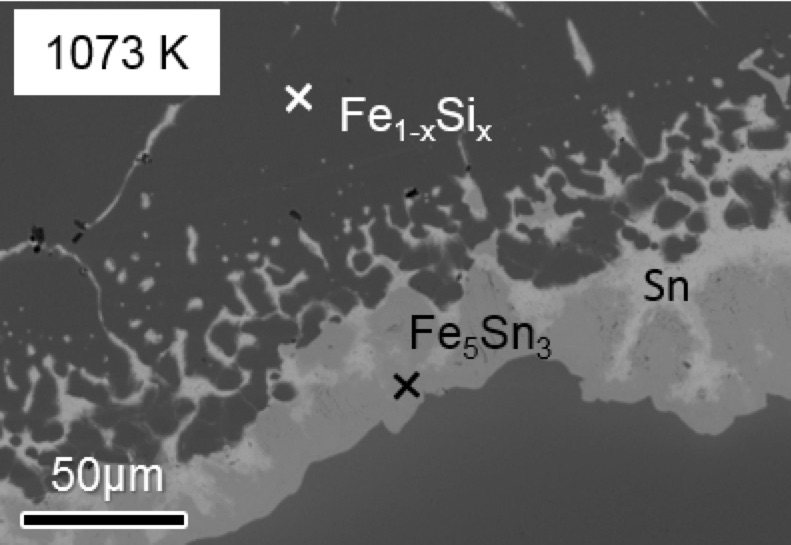}
\caption{\label{fig:bahar_s5} BSE image of the diffusion zone in the Fe-Sn$_{0.5}$Si$_{0.5}$ crucibles annealed at 1073~K. The binary Fe$_3$Sn phase is not formed in the crucible annealed at 1073~K.}
\end{figure}

\begin{figure}[tbp]
\centering
\includegraphics[scale=0.3]{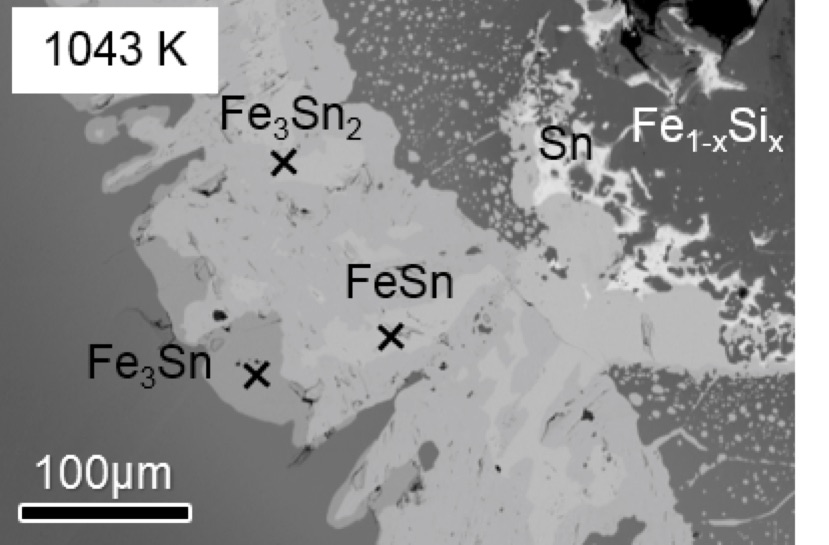}
\caption{\label{fig:bahar_s4} BSE image of the diffusion zone in the Fe-Sn$_{0.5}$Si$_{0.5}$ crucibles annealed at 1043~K. The binary Fe$_3$Sn phase is formed in the crucible annealed at 1043~K. However, Si totally dissolved in Fe in a separate region and did not diffuse into the 3:1 structure.}
\end{figure}

Figures \ref{fig:bahar_s4} and \ref{fig:bahar_s5} show the diffusion zone of the Fe-Sn$_{0.5}$Si$_{0.5}$ crucibles annealed at 1073~K and 1043~K respectively. In the crucible annealed at 1073~K, the high temperature phase Fe$_5$Sn$_3$ was formed whereas at lower temperature (1043~K), FeSn, Fe$_3$Sn$_2$ and Fe$_3$Sn phases were formed. Although the 3:1 structure is formed in the crucibles, however, according to the EDX quantification, Si did not diffuse into 3:1 structure.

%\end{document}

\newpage
\clearpage

% use "aiaa" style for BibTeX

%\section*{References}

\end{document}